\documentclass[10pt,journal,compsoc]{IEEEtran}
\usepackage[nocompress]{cite}

\usepackage{amsmath}

\ifCLASSOPTIONcompsoc
  \usepackage[caption=false,font=footnotesize,labelfont=sf,textfont=sf]{subfig}
\else
  \usepackage[caption=false,font=footnotesize]{subfig}
\fi

\usepackage{stfloats}

\usepackage[utf8]{inputenc} 
\usepackage[T1]{fontenc}
\usepackage{url}
\usepackage{ifthen}

 \usepackage[usenames,dvipsnames]{pstricks}
 \usepackage{epsfig}
 \usepackage{pst-grad} 
 \usepackage{pst-plot} 
 
\usepackage{graphicx}
  \usepackage{amsthm}
  \usepackage{mathdots}
  \usepackage{amssymb,bm}
  \usepackage{amsfonts}
  \usepackage{amscd}
  \usepackage{color,xcolor}
  \usepackage{epsfig}
\usepackage{psfrag}
\usepackage{xfrac}
  \usepackage{dsfont}
  \usepackage{textcomp}
  \usepackage{etoolbox}
\usepackage{verbatim} 
\usepackage{lettrine}
   \usepackage{algorithm}
   \usepackage{algorithmicx}
   \usepackage{algpseudocode}
  \usepackage{enumerate} 
  \usepackage[font={it,rm}]{caption} 
\usepackage{soul}
\setstcolor{blue}
\usepackage{url}
\usepackage{tikz}
\usetikzlibrary{arrows,decorations.markings}
\usetikzlibrary{decorations.pathreplacing,angles,quotes,shapes}

\definecolor{myblue}{rgb}{0,0,0.5}

\newtheorem{lemma}{Lemma}
\newtheorem{theorem}{Theorem}

\newtheorem{definition}{Definition}

\newtheorem{remark}{Remark}  

\newtheorem{property}{Property}  

\newcommand{\salpha}{\scalebox{0.97}{$\frac{p}{1-p}$}}
\newcommand{\salphainv}{\scalebox{0.97}{$\frac{1-p}{p}$}}

\newtoggle{SINGLE_COL}
\toggletrue{SINGLE_COL}
\togglefalse{SINGLE_COL}

\usepackage{setspace}
\AtBeginEnvironment{algorithm}{\singlespacing}

\IEEEoverridecommandlockouts 

\begin{document}
\title{Impact of Fake Agents on Information Cascades} 

\author{Pawan~Poojary,~\IEEEmembership{Member,~IEEE,}
        and~Randall~Berry,~\IEEEmembership{Fellow,~IEEE}
\IEEEcompsocitemizethanks{\IEEEcompsocthanksitem P. Poojary and R. Berry are with the Department
of Electrical and Computer Engineering, Northwestern University, Evanston, IL - 60208.\protect\\
E-mail: \{pawanpoojary2018@u, rberry@ece\}.northwestern.edu
\IEEEcompsocthanksitem This paper was presented in part at the IEEE International Symposium on Information Theory, Los Angeles, CA, USA, June 2020 \cite{poojary2020observational}.
\IEEEcompsocthanksitem This work was supported in part by the NSF under grants CNS-1701921, CNS-1908807, ECCS-2030251 and ECCS-2216970.}
}

\IEEEtitleabstractindextext{%
\begin{abstract}
In online markets, agents often learn from other's actions in addition to their private information. Such observational learning can lead to \emph{herding} or \emph{information cascades} in which agents eventually ignore their private information and ``follow the crowd''. Models for such cascades have been well studied for Bayes-rational agents that arrive sequentially and choose pay-off optimal actions. This paper additionally considers the presence of \emph{fake agents} that take a fixed action in order to influence subsequent rational agents towards their preferred action. We characterize how the fraction of such fake agents impacts the behavior of rational agents given a fixed quality of private information. Our model results in a Markov chain with a countably infinite state space, for which we give an iterative method to compute an agent's chances of herding and its welfare (expected pay-off). Our main result shows a counter-intuitive phenomenon: there exist infinitely many scenarios where an increase in the fraction of fake agents in fact reduces the chances of their preferred outcome. Moreover, this increase causes a significant improvement in the welfare of every rational agent. Hence, this increase is not only counter-productive for the fake agents but is also beneficial to the rational agents.
\end{abstract}

}

\maketitle

\IEEEpeerreviewmaketitle

\IEEEraisesectionheading{\section{Introduction}\label{sec:intro}}
\IEEEPARstart{C}{onsider} a new item that is up for sale in a recommendation-based market where agents arrive sequentially and decide whether to buy the item, with their choice serving as a recommendation for later agents (eg., through a public database). This item has a common quality/utility (say ``good'' or ``bad'')  that is unknown to the agents. Each agent then makes a pay-off optimal decision by using its own prior knowledge of the item's quality and by observing the choices of its predecessors. Such models of ``observational learning'' were first studied by \cite{bhw,banerjee,welch} under a Bayesian learning framework wherein each agent has some prior knowledge in the form of a privately observed signal about a pay-off-relevant state of the world generated from a commonly known probability distribution. Agents arrive in an exogenous order, and every agent chooses its action based on its own private signal and observations of past agents' actions. Agents are assumed to be Bayes \emph{rational}, i.e., their actions are optimal with repsect to their posterior beliefs of the true state of the world given their observations\footnote{From a game theoretic point-of-view, this is a dynamic game with assymmetric information where the agents' optimal actions are a Perfect Bayesian Equilibrium (PBE) of this game.}. A key result for such models is the emergence of \emph{information cascades} or \emph{herding}, i.e., at some point, it is optimal for an agent to ignore its own private signal and follow the actions of the past agents. Subsequent agents then follow suit due to their homogeneity. As a result, from the onset of a cascade, the agents' actions do not reveal any information conveyed to them by their private signals; hence learning stops.



The model described above has two possible outcomes. First, the agents may end up in a \emph{correct} cascade, that is the information accumulated within the observation history eventually forces all successive agents to take the optimal action with respect to the underlying true state of the world. In this case, \emph{learning} or a socially optimal behaviour is said to be achieved. Second, there is also the possibility that past observations might get accumulated in a manner that forces all successive agents to take the action that, although individually optimal, is in fact socially sub-optimal, which we refer to as a \emph{wrong} cascade. Thus, herding to a wrong cascade would prevent the agents from learning the socially optimal (correct) choice. Moreover, a main result of work in \cite{bhw,banerjee,welch} (and others) that assumes homogeneous agents and discrete bounded private signals is that a cascade (correct or wrong) commences within the arrival of a finite number of agents with probability one. This leads to a positive probability that a wrong cascade occurs.

In this paper, we study a Bayesian learning model similar to \cite{bhw,banerjee,welch} where in addition to rational agents, we introduce randomly arriving \emph{fake} agents that always take (or fake) a fixed action, regardless of their pay-off, in order to influence the outcome of a cascade. For example, this could model a setting where a poor quality item is posted for sale on a website that records buyers' decisions, and fake agents intentionally buy (or appear to buy) this item to make it seem worth buying to future buyers \cite{forster2016trust}. The model could also represent a setting where rational actions are manipulated or where fake actions are inserted into a recommendation system \cite{hu2012manipulation}, \cite{akoz2017manipulation}, \cite{harvard_review}. The objective is to study the impact of varying the amount of these fake agents on the probability of their preferred cascade and on the resulting \textit{welfares} (expected pay-offs) of the rational agents. Our main result shows a counter-intuitive phenomenon: the probability with which the cascade preferred by the fake agents occurs is not monotonically increasing in the fraction of fake agents, $\epsilon$. In fact, there exist infinitely many cases where an increase in the fraction of fake agents  reduces the chances of their preferred cascade. We identify a sequence of thresholds for $\epsilon$ where this phenomenon is most pronounced. Moreover, we show that exceeding any such $\epsilon$ threshold causes an abrupt improvement in the welfare of every rational agent.  



In our model, we assume that the presence of fake agents is common knowledge to the buyers and is accounted for in their actions. This is motivated by several empirical results, such as in \cite{kumar2024exploring} and \cite{huang2023unmasking}, which suggest that over repeated interactions with the buying platform and through word-of-mouth, buyers tend to be conservative with their decisions by taking into account that a certain fraction of positive reviews for an item might be fake or that its sales statistics might be exaggerated.


 




\subsection{Our Contributions}



In this paper, we develop a Markov chain model to represent the process of information accumulation driven by the actions of sequentially arriving agents (either rational or fake), which eventually results in a cascade. To obtain this Markov chain, we identify a sufficient statistic of past observations that determines whether an agent cascades or follows its private signal.

Secondly, we analyse the Markov chain for the probability of cascades as a function of the quality $p$ of the prior knowledge available to every rational agent (i.e., its private signal), and the fraction of fake agents, $\epsilon$. The Markov chain typically occupies a countably infinite state-space, and does not readily allow for a closed-form solution to the cascade probabilities. Instead, we develop recursive equations that can compute the cascade probabilities with arbitrary precision. These equations are motivated by the construction of an iterative method that enumerates all possible action sequences that would lead to a cascade. This iterative method also provides exact probabilities for both correct and wrong cascades to begin within the arrival of a finite number of agents. Computing these cascade probabilities in turn yields the welfare for any arbitrary agent. 

Thirdly, we study the effects of varying $\epsilon$ on the cascade probabilities and on agents' welfares for a fixed private signal quality,\, $p$. Our results highlight the following counter-intuitive phenomenon: there exist an infinite sequence of thresholds $\{\epsilon_r\}_{r=1}^\infty$ where increasing $\epsilon$ slightly above $\epsilon_r$ causes an abrupt decrease in the probability of the cascade preferred by the fake agents. Thus, a marginal increase in the presence of fake agents beyond any such threshold reduces the chances of their preferred cascade instead of increasing it, which is what they had intended. Further, we analytically show that increasing $\epsilon$ just beyond any such $\epsilon_r$ also causes an abrupt and significant increase in the welfares of all agents. Therefore, marginally exceeding $\epsilon_r$ is not only counter-productive for the fake agents but is also beneficial to the rational agents.


Fourthly, we prove that an agent's welfare improves if it chooses to arrive later in the sequence of arrivals. This result implies that the welfare reaches a limiting maximum value as the agent's arrival index tends to infinity. We refer to this value as the long run or \emph{asymptotic} welfare. The proof involves the application of Blackwell's theorem on comparing information structures\cite{leshno1992elementary}, \cite{blackwell1953}, which more generally also shows that an agent can never do any better by ignoring any of the past observations or its own signal.


Fifthly, we quantify the cascade probabilities and social welfare as a function of $p$ in the interesting scenario where the proportion of fake agents approaches unity. We analytically show that even when fake agents have overwhelmed the rational agents: $(i)$ a better signal quality results in higher chances of learning the socially optimal action (correct cascade) and $(ii)$ rational agents continue to benefit from past observations.



Sixthly, we discuss the role of a \emph{Platform Co-ordinator} in improving learning through modifications to  the observation database. We analyse and compare the performances of three approaches for improving the agents’ welfares, namely $(a)$ increasing the fraction of fake actions, $(b)$ filtering out possibly fake actions and $(c)$ modifying the possibly fake actions. For a wide range of values for $p$, we observe that for low values of $\epsilon$, Scenario $(c)$ outperforms all other scenarios, providing the best improvement in welfare. As $\epsilon \rightarrow 0$, it entirely mitigates the reduction in welfare caused by the fake agents. Whereas, for high values of $\epsilon$, Scenario $(b)$ provides the best improvement. There also exist several intervals with moderate $\epsilon$-values where Scenario $(a)$ performs the best.


We conclude by showing that the analysis, results and discussions presented in this work readily extend to general priors and agent pay-offs, as long as the ex-ante pay-off is zero. Moreover, our analytic techniques can be easily modified to include a non-zero ex-ante pay-off.





\subsection{Related Work}   
Our work follows the basic model studied in \cite{bhw,banerjee,welch}, except we introduce fake agents which arrive at random amidst the sequence of rational agents. Many variations of this basic model have been studied, some of which we review here briefly. The work in \cite{smithNsorensen} relaxed the assumptions of agents' homegeneity and binary valued private signals made in \cite{bhw,banerjee,welch}. It showed in particular that allowing for a richer class of signals, such that their likelihood ratio is unbounded, could result in learning to occur with probability one. Our model maintains the assumptions of \cite{bhw,banerjee,welch}, i.e., homegenous agents and binary valued private signals.

Another change to the basic model is to consider different structures for observing past actions. For example, 
\cite{acemouglu} considers that agents can observe only a subset of the past actions, defined by an underlying network structure. This work finds conditions on the network structure that guarantee asymptotic learning. Whereas \cite{song} makes the network formation endogenous by allowing agents to select their observations at a cost. Another variation, studied in \cite{sgroi2002optimizing}, forces a fixed number of initial agents to only observe their private signals. These agents thereby act as "guinea pigs", used to explore the unknown true value. This causes improved welfares of the subsequent agents at the cost of the guinea pigs’ welfares. Our work here stays with the original model \cite{bhw,banerjee,welch} which assumes that actions are recorded to a common database, thereby allowing each agent to observe all prior actions (although in our case, each action could be either rational or fake).


Closer to our work is the model in \cite{Tho}, which assumes that the recording of actions for subsequent agents is subject to an error that is unbiased towards each of the possible actions. In our setting, an action being either fake or rational depending on the agent-type could equivalently be perceived as an error while recording a rational action (as in \cite{Tho}); except that in our case, the error is biased only towards a preferred action.\footnote{This change requires a different analysis approach than that in \cite{Tho} as the underlying Markov chain now typically has a countably infinite state-space, while in \cite{Tho} it was finite. We show that this change also yields substantially different outcomes.} 

There is also a body of work that considers agents similar to our fake agents, who only take a single action regardless of the true state. This includes the \textit{crazy agents} considered in \cite{smithNsorensen}, \emph{stubborn agents} in \cite{acemouglu2013opinion} and \emph{zealots} in \cite{mobilia2003zealot}. While \cite{smithNsorensen} relaxes the assumption of binary signals, it does not consider how changing the fraction of crazy agents affects the cascade probability, which is the main focus of our work. The works in \cite{acemouglu2013opinion} and \cite{mobilia2003zealot} do not consider learning, instead they model opinion/belief dynamics in the presence of their respective types of abnormal agents. They consider non-Bayesian models for updating agents' beliefs, while our work remains with the Bayesian model. Other types of agents considered in the literature include the \emph{revealers} in \cite{peres2017}, \emph{experts} and \emph{laymen} in \cite{Wu2015} and \emph{non-myopic} agents in \cite{Achilleas2022}.

The learning models considered in this line of work also have ties to early work on sequential detection with finite memory, e.g, \cite{cover1969hypothesis} and \cite{hellman1970learning}. There, the sequence of i.i.d. signals has to be summarized by a test-statistic of finite cardinality, which gets updated as per a rule designed by a planner with the objective of maximizing the chance of asymptotically learning the true state of the world. Our model does not assume any memory constraints and all past actions are perfectly observed. The more important distinction is that in \cite{cover1969hypothesis,hellman1970learning}, agents simply follow the rule prescribed by the planner, i.e., agents are not strategic and thus their actions might not constitute an equilibrium. Our work instead considers a setting where there is no planner and agents are strategic and act in their best interests. Thus, agents' actions are in a PBE. 

Our work is a substantial extension of \cite{poojary2020observational}, where we first proposed the iterative method to compute cascade probabilities and identified the infinite sequence of $\epsilon$-thresholds at which an abrupt reduction in the preferred cascade probability occurs. This paper makes several contributions beyond \cite{poojary2020observational}, such as studying the cascade probability for finite agents, identifying important properties exhibited by agents' welfares, and exploring the effects of a Platform Co-ordinator on learning, among others as stated earlier.

\vspace{-2.5mm}
\subsection{Organisation}
The remainder of the paper is organized as follows. We describe our model in Section \ref{sec:model}. We analyze this model and identify the resulting cascade properties in Section \ref{optimal_and_cascades}. In Section \ref{sec:markov}, we present our Markov chain formulation, identify error thresholds and devise an iterative method to compute cascade probabilties. Further, in Subsection \ref{subsec:abrupt_drop_Pycas}, we quantify the abrupt reduction in chances of a correct cascade at these error thresholds. Subsection \ref{subsec:PYcas_low_epsilon} investigates the preferred cascade probability for low values of $\epsilon$ and contrasts it to the case when fake agents are absent $(\epsilon=0)$. Section \ref{sec:Exp_pay-off} characterizes agents' welfares and identifies important properties exhibited by them. In Section \ref{sec:learning_limeps_1}, we investigate learning in the limiting scenario where the proportion of fake agents approaches one. Section \ref{sec:platform_coord} introduces a Platform Co-ordinator and presents approaches for modifying the observations that could improve learning. Section \ref{sec:general_prior} extends this work to general priors and agent pay-offs, while maintaining zero ex-ante pay-off. Lastly, we present our conclusions in Section \ref{concl}. Detailed proofs, extended analyses and supporting plots are provided in the Appendix.

\section{Model} 
\label{sec:model}


We consider a model similar to \cite{bhw} in which there is a countable sequence of agents, indexed $i = 1,2, \ldots$ where the index represents both the time and the order of actions. Each agent $i$ takes an action $A_i$ of either buying $(Y)$ or not buying $(N)$ a new item that has a true value $(V)$, which could either be good $(G)$ or bad $(B)$. For simplicity, both possibilities of $V$ are assumed to be equally likely.

The agents are Bayes-rational utility maximizers where the pay-off received by each agent $i$, denoted by $\pi_i$, depends on its action $A_i$ and the true value $V$ as follows. If the agent chooses $N$, his payoff is $0$. Whereas, if the agent chooses $Y$, he incurs a cost of $C = 1/2$ for buying the item and gains an amount that reflects the item's value/utility to its buyer. The buyer loses an amount $y=0$ if $V = B$ and gains an amount $x=1$ if $V = G$. The agent's net pay-off is given by
\begin{align}
\pi_i = \begin{cases}   \text{\scalebox{0.93}{$x-C = 1/2,$}}  &\text{if} \; \text{\scalebox{0.95}{$A_i=Y$}} \; \text{and} \; \text{\scalebox{0.95}{$V=G,$}} \\
\text{\scalebox{0.93}{$-y-C = -1/2,$}}  &\text{if} \; \text{\scalebox{0.95}{$A_i=Y$}} \; \text{and} \; \text{\scalebox{0.95}{$V=B,$}} \\
0,  &\text{if} \; \text{\scalebox{0.95}{$A_i=N.$}} \\
\end{cases}
\end{align}
Given the values considered for $x,y$ and $C$, observe that since $V$ is equiprobable, the \emph{ex ante} expected pay-off for any agent is $0$ for either of the actions.\footnote{Section \ref{sec:general_prior} generalizes the model to possibly non-uniform priors for $V$ and a general pay-off structure for agents, while still retaining the condition of zero ex-ante pay-off.} Thus, to begin with, an agent is indifferent to the two actions. 







\begin{figure}[h]
\centering
\subfloat[]{
\centering
\begin{tikzpicture}[scale=1.4]
\draw [decoration={markings,mark=at position 1 with {\arrow[scale=2,>=stealth]{>}}},postaction={decorate}] (0,0) -- (1.5,0);
\draw [decoration={markings,mark=at position 1 with {\arrow[scale=2,>=stealth]{>}}},postaction={decorate}] (0,-1) -- (1.5,-1);
\draw [decoration={markings,mark=at position 1 with {\arrow[scale=2,>=stealth]{>}}},postaction={decorate}] (0,0) -- (1.5,-1);
\draw [decoration={markings,mark=at position 1 with {\arrow[scale=2,>=stealth]{>}}},postaction={decorate}] (0,-1) -- (1.5,0);
\node at (1.65,0) {$H$};
\node at (1.65,-1) {$L$};
\node at (-0.15,0) {$G$};
\node at (-0.15,-1) {$B$};
\node at (-0.35,-0.5) {$V$};
\node at (1.85,-0.5) {$S_i$};
\node at (0.75,0.15) {\scalebox{0.83}{$p$}};
\node at (0.75,-1.15) {\scalebox{0.83}{$p$}};
\node at (0.54,-0.22) [rotate= -33.7] {\scalebox{0.83}{$1-p$}};
\node at (0.56,-0.8) [rotate= 33.7] {\scalebox{0.83}{$1-p$}};
\end{tikzpicture}
\label{BSC}}
\hspace{0.05\linewidth}
\subfloat[]{
\centering
\begin{tikzpicture}[scale=1.4]
\draw [decoration={markings,mark=at position 1 with {\arrow[scale=2,>=stealth]{>}}},postaction={decorate}] (0,0) -- (1.5,0);
\draw [decoration={markings,mark=at position 1 with {\arrow[scale=2,>=stealth]{>}}},postaction={decorate}] (0,-1) -- (1.5,-1);
\draw [decoration={markings,mark=at position 1 with {\arrow[scale=2,>=stealth]{>}}},postaction={decorate}] (0,-1) -- (1.5,0);
\node at (1.65,0) {$Y$};
\node at (1.65,-1) {$N$};
\node at (-0.15,0) {$Y$};
\node at (-0.15,-1) {$N$};
\node at (-0.35,-0.5) {$A_i$};
\node at (1.85,-0.5) {$O_i$};
\node at (0.75,0.15) {\scalebox{0.9}{$1$}};
\node at (0.75,-1.15) {\scalebox{0.9}{$1-\epsilon$}};
\node at (0.8,-0.6) [rotate= 33.7] {\scalebox{0.9}{$\epsilon$}};
\end{tikzpicture}
\label{ASC}}
\setlength{\abovecaptionskip}{8pt}
\setlength{\belowcaptionskip}{-1pt}
\caption{\small $(a)$ The BSC through which agents receive private signals. \; $(b)$ The channel through which agents' actions are corrupted.}
\end{figure}

To incorporate agents' private beliefs about the new item, every agent $i$ receives a private signal $S_i \in \{ H \, \text{(high)}, L \, \text{(low)}\}$. This signal, as shown in Figure \ref{BSC}, partially reveals the information about the true value of the item through a binary symmetric channel (BSC) with crossover probability $1-p$, where $\sfrac{1}{2} < p < 1$. This implies that the signal is informative but not revealing. Moreover, the sequence of private signals $\{S_1,S_2,\ldots\}$ is assumed to be \emph{i.i.d.} given the true value $V$. Each agent $i$ takes a \emph{rational} action $A_i$ that depends on his private signal $S_i$ and the past observations $\{O_1,O_2, \ldots, O_{i-1} \}$ of actions $\{A_1,A_2, \ldots, A_{i-1} \}$. Next, we modify the information structure in \cite{bhw} by assuming that at each time instant, an agent could either be \emph{fake} with probability (w.p.) $\epsilon \in [0,1)$ or \emph{ordinary} w.p. $1-\epsilon$, where the value $\epsilon$ is assumed to be common knowledge, so that all agents know the probability that any agent is fake but do not know which specific agents are fake. An ordinary agent $i$ \emph{honestly} reports his action, \emph{i.e.} $O_i=A_i$. On the contrary, a fake agent \emph{always} reports a $Y$, reflecting his intention of influencing the successors into buying the new item, regardless of its true value. This implies that at any time $i$, if $A_i = N$ then with probability $ 1-\epsilon $, the reported action $O_i = N$ and with probability $\epsilon$, $O_i = Y.$ Whereas, if $A_i = Y$ then $O_i = Y$ with probability $1$. Refer to Figure \ref{ASC}. 

An equivalent model is where action $A_i$ is rational only if agent $i$ is ordinary and is fixed to $Y$ otherwise, while $O_i = A_i$ for all agents. This yields the same information structure as the model above and so the same analysis applies to model the behavior of the ordinary agents. We chose the former model mainly to simplify our notation.

\section{Optimal decision, cascades and welfare} 
\label{optimal_and_cascades}
For the $n^{\text{th}}$ agent, let the history of past observations be denoted by $\mathcal{H}_{n-1} \! = \! \{O_1,O_2, \ldots, O_{n-1} \}$. As the first agent does not have any observation history, he always follows his private signal, i.e., he buys if and only if the signal is $H$. From the second agent onwards, the Bayes' optimal action for every agent $n$, $A_n$ is chosen according to the hypothesis ($V \! =G$ or $B$) that has the higher posterior probability given the information set $\mathbb{I}_n = \{S_n, \mathcal{H}_{n-1} \}$. Let $\gamma_n (S_n, \mathcal{H}_{n-1} ) \triangleq \mathbb{P} (G \vert S_n, \mathcal{H}_{n-1} )$ denote the posterior probability for the item being good, \scalebox{0.98}{$V \! = G$}. Then the Bayes' optimal decision rule is
\begin{align}
A_n = \begin{cases} 
Y, \quad & \text{if} \; \; \gamma_n  > 1 / 2, \\
N, \quad & \text{if} \; \; \gamma_n  < 1 / 2, \\
\text{follows}\;\; S_n,  \quad & \text{if} \; \; \gamma_n = 1 / 2.
\end{cases} 
\label{bayes_decison}
\end{align}
Note that when $\gamma_n \!=\! 1/2$, an agent is indifferent to the two actions. Similar to \cite{Tho}, our decision rule in this case follows the private signal $S_n$, unlike \cite{bhw}, which employs a randomized tie-breaking rule. Another choice in this case is to follow the history \scalebox{0.95}{$\mathcal{H}_{n-1}$,} i.e., to take the action that is most observed in the past. Techniques in this paper can be readily adapted to reflect this alternate choice of breaking ties \cite{pawanWiopt2023}.

\begin{definition}
An information cascade is said to occur when an agent's decision becomes independent of his private signal.
\end{definition}

It follows from \eqref{bayes_decison} that, agent $n$ cascades to a $Y$ $(N)$ if and only if $\gamma_n  > 1 / 2 $ $( < 1/2)$ for all $S_n \in \{ H,L \}$. The other case being $\gamma_n  \geq 1 / 2$ for $S_n=H$ and $\gamma_n  \leq 1 / 2$ for $S_n=L$; in which case, agent $n$ follows $S_n$. A more intuitive way to present this condition is to first express the information contained in the history $\mathcal{H}_{n-1}$ observed by agent $n$ in the form of its public likelihood ratio, 
\begin{align}
l_{n-1} (\mathcal{H}_{n-1}) := \frac{\mathbb{P}(\mathcal{H}_{n-1} \vert B) }{ \mathbb{P}(\mathcal{H}_{n-1} \vert G)},
\end{align}
and then state it as follows. 
\begin{lemma}
Agent $n$ cascades to a $Y \, (N)$ if and only if $l_{n-1} < \frac{1-p}{p} $ $ \big( l_{n-1} > \frac{p}{1-p} \big)$ and otherwise follows its private signal $S_n$.
\label{lemma1}
\end{lemma}
\vspace{1mm}

To prove this lemma, first define agent $n$'s private likelihood ratio, $\beta_n(S_n) := \mathbb{P}(S_n \vert B) / \mathbb{P}(S_n \vert G)$. It follows from Figure \ref{BSC} that $\beta_n(H) = (1-p)/p$ and $\beta_n(L) = p/(1-p)$. Next, using Bayes' rule, express $\gamma_n$ in terms of $l_{n-1}$ and $\beta_n$ as $\gamma_n = 1/(1+\beta_n l_{n-1})$. As a result, the condition on $\gamma_n$ for a $Y \, (N)$ cascade translates to $l_{n-1} <1 / \beta_n$ $(>1 / \beta_n)$ for all $S_n$; this simplifies to give Lemma \ref{lemma1}.

If agent $n$ cascades, then the observation $O_n $ does not provide any additional information about the true value $V$ to the successors over what is contained in $\mathcal{H}_{n-1}$. As a result, $l_{n+i} = l_{n-1} $ for all $i=0,1,2,\ldots$ and hence they remain in the cascade, which leads us to the following property, also exhibited by prior models, e.g. \scalebox{1}{\cite{bhw,banerjee,welch,pawanMFG,pawan_congest,more_choices2024}.}
\begin{property}
Once a cascade occurs, it lasts forever.
\label{prop1}
\end{property}
\noindent On the other hand, if agent $n$ does not cascade, then Property \ref{prop1} and Lemma \ref{lemma1} imply that all the agents until and including $n$ follow their own private signals ignoring the observations of their predecessors. For every such observation $O_i$, $i \leq n$, as $S_i$ is conditionally independent of the history $\mathcal{H}_{i-1}$ given $V,$ the likelihood ratio can be updated as 
\begin{align}
l_i = \begin{cases} \Big( \frac{1-b}{a} \Big) l_{i-1}, \quad & \text{if} \;\; O_i = Y, \vspace{0.8mm} \\
\Big( \frac{b}{1-a} \Big) l_{i-1},  \quad & \text{if} \;\; O_i = N,
\end{cases}  \label{likelihood_update}
\end{align}
\begin{equation}
a := \mathbb{P} (O_i = Y \vert V=G) \;\; \text{and} \;\; b := \mathbb{P} (O_i = N \vert V=B).
\label{a_b_definitions}
\end{equation} 
Here, $a$ and $b$ denote the probabilities that an observation $O_i$ follows $V$ if agent $i$ follows its private signal, given $V=G$ and $B$, respectively. It can be shown from Figures \ref{BSC} and \ref{ASC} that in the above case, \emph{i.e.}, when $A_i$ follows $S_i$,
\begin{align}
a = p+(1-p) \epsilon \;\; \text{and} \;\; b = p (1-\epsilon).
\label{a_b_expressions}
\end{align}

As a result of the updates, $l_n$ can be shown to depend only on the number of $Y$'s (denoted by $n_Y$) and $N$'s (denoted by $n_N$) in the observation history $\mathcal{H}_{n}$. Specifically, \scalebox{0.98}{$l_n = \left( \frac{1-p}{p} \right)^{h_n}$} where $h_n$ is the difference between the number of $Y$'s weighted by $\eta$ and the number of $N$'s,
\begin{align}
h_n =& \; \eta n_Y - n_N, \label{h_n}  \\
\eta :=& \; \log \Big( {\small \text{$\frac{a}{1-b}$}} \Big) / \log \Big( {\small \text{$ \frac{p}{1-p}$}} \Big).  \label{eta}
\end{align} 
Thus, agents that have not yet entered a cascade satisfy the following property.
\begin{property}
Until a cascade occurs, each agent follows its private signal. Moreover, $h_n$ defined in \eqref{h_n} is a sufficient statistic of the information contained in the past observations.
\label{prop2}
\end{property}

Note that if $\epsilon = 0$ (no fake agents) then $a = b = p$ and $\eta = 1$, in which case $h_n$ is the {\it unweighted} difference, $\eta_Y- \eta_N$, which is also the case for the unbiased noise model in \cite{Tho}. Whereas, if $\epsilon >0$ then $\eta<1$. The expression for $h_n$ in \eqref{h_n} shows that, due to the presence of fake agents, the dependence of an agent’s decision on a $Y$ in his observation history reduces by a factor of $\eta$, whereas the dependence on a $N$ remains unaffected. This is to be expected because, unlike a $N$ which surely comes from an honest agent, a $Y$ incurs the possibility that the agent could be fake. Further, this reduced dependence on $Y$ is exacerbated with an increase in the possibility of fake agents, as $\eta$ reduces with an increase in $\epsilon$.

Using the expression for $l_n$ in Lemma \ref{lemma1}, it follows that for all times $n$ until a cascade occurs, $-1 \leq h_n \leq 1$ and the update rule for $h_n$ is given by
\begin{align}
h_n = \begin{cases} h_{n-1} + \eta, \quad & \text{if} \;\; O_n = Y,\\
h_{n-1}-1,  \quad & \text{if} \;\; O_n = N.
\end{cases}
\label{mc_update}
\end{align}
Whereas, once $h_n >1 $ ($ < -1 $), a $Y \,(N)$ cascade begins and $h_n$ stops updating (Property \ref{prop1}). Note that $h_0 = 0$ since the first agent has no observation history. Now, given the true value $V \in \{G,B\}$, let the probability that a $Y$ $(N)$ cascade begins be denoted by $\mathbb{P}^V_{Y\text{-cas}}$ $\big( \mathbb{P}^V_{N\text{-cas}} \big)$. Here, $ \mathbb{P}^V_{N\text{-cas}} = 1- \mathbb{P}^V_{Y\text{-cas}}$ as it can be shown that the process $\{h_n\}$ exits the range $[-1,1]$ w.p. $1$. Further, let the $n^{\text{th}}$ agent's \emph{welfare} refer to its pay-off averaged (in expectation) over $V \in \{G,B\}$. We show later in Section \ref{sec:Exp_pay-off} that this welfare as $n \rightarrow \infty$ relates to the cascade probabilities of the process $\{h_n\}$ as 
\begin{align}
\Pi &:= \underset{n \rightarrow \infty}{\lim} \mathbb{E} [\pi_n ]  = \frac{1}{4} \left[ \mathbb{P}_{Y\text{-cas}}^G  - \mathbb{P}_{Y\text{-cas}}^B \right].
\label{Pi_expression}
\end{align} 


\section{Markovian Analysis of Cascades} \label{sec:markov}
In this section, we analyse the process $\{h_n\}$, given $V,$ to determine the probability of cascades. It follows from the previous section that conditioned on $V$, the process $\{ h_n\}$ is a discrete-time Markov chain taking values in $[-1,1]$ before getting absorbed into the \emph{left absorption region} $(< -1)$ causing a $N$ cascade or the \emph{right absorption region} $( > 1)$ causing a $Y$ cascade. More specifically, equation \eqref{mc_update} shows that, given $V,$ $ \{ h_n \}$ is a random walk (r.w.) that starts from state $0$ and moves to the right by $\eta$ w.p. $\mathbb{P}(O_n = Y \vert V)$ or to the left by $1$ w.p. $\mathbb{P}(O_n = N \vert V) $ until a cascade occurs, where these probabilities are defined in terms of $a$ and $b$ in \eqref{a_b_definitions}. Figure \ref{random_walk} depicts this random walk, where $p_f \triangleq \mathbb{P}(O_n = Y \vert V)$ denotes the probability of a $Y$ being observed given $V$, when any agent $n$ follows its private signal $S_n$. We have from \eqref{a_b_definitions} that $p_f=a$ for $V=G$, whereas $p_f= 1-b$ for $V=B$.



Note that in the special case where $\eta$, given by \eqref{eta}, satisfies $1 / \eta = r$ for some $r=1,2,\ldots$, the process $\{ h_n\}$ is equivalent to a Markov chain with finite state-space \scalebox{0.9}{$\mathcal{A} = \{ -r-1,-r, \ldots ,-1,0,1,\ldots,r,r+1 \} $}, and with $-r-1$ and $r+1$ being absorption states corresponding to $N$ and $Y$ cascades, respectively. More generally, it can be proved that $\{h_n\}$ has a finite state-space for any rational-valued $1 / \eta$. In such cases, absorption probabilities can be obtained directly by solving a system of linear equations. In this paper, our main focus is on the more generic case of irrational values of $1 / \eta$ resulting in $\{ h_n\}$ taking countably infinite values in $[-1,1],$\footnote{For example, if $\eta$ was chosen uniformly at random, then almost surely (w.p. $\! 1$) it would fall into this case.} which does not readily allow for a direct solution to the cascade probabilities. This is unlike the unbiased noise model in \cite{Tho} where the state-space of the Markov chain is always finite. 



\begin{figure}[h!]
\centering
\begin{tikzpicture}[scale=1.8]
\draw [decoration={markings,mark=at position 1 with {\arrow[scale=2,>=stealth]{>}}},postaction={decorate}] (0,0) -- (3.5,0);
\draw [decoration={markings,mark=at position 1 with {\arrow[scale=2,>=stealth]{>}}},postaction={decorate}] (3.5,0) -- (0,0);
\draw (0.2,0.05) -- (0.2,-0.05);
\draw (1.75,0.05) -- (1.75,-0.05);
\draw (3.3,0.05) -- (3.3,-0.05);
\node at (1.8,-0.15) {\scalebox{0.9}{$0$}};
\draw [->] (1.75,0) to [out=90,in=90] (2.15,0); \node at (1.97,0.22) {\scalebox{0.9}{$p_f$}}; \node at (2.2,-0.15) {\scalebox{0.9}{$\eta$}};
\draw [->] (2.15,0) to [out=90,in=90] (2.55,0); \node at (2.37,0.22) {\scalebox{0.9}{$p_f$}}; \node at (2.55,-0.15) {\scalebox{0.9}{$2\eta$}};
\draw [->] (1.75,0) to [out=-90,in=-90] (0.2,0); \node at (0.87,-0.6) {\scalebox{0.9}{$1-p_f$}};
\node at (0.10,-0.15) {\scalebox{0.9}{$-1$}};
\draw [->] (2.15,0) to [out=-90,in=-90] (0.6,0); \node at (1.48,-0.6) {\scalebox{0.9}{$1-p_f$}};
\node at (0.64,-0.16) {\scalebox{0.9}{$-1+ \eta$}};
\node at (3.3,-0.15) {\scalebox{0.9}{$1$}};
\draw [dashed,->] (2.55,0) to [out=90,in=90] (2.95,0);
\draw [dashed,->] (2.55,0) to [out=-90,in=-90] (1.0,0);
\end{tikzpicture}
\setlength{\abovecaptionskip}{3pt}
\setlength{\belowcaptionskip}{-5pt}
\caption{\small Partial transition diagram of random walk $\{h_n\}$ given $V$.}  
\label{random_walk}
\end{figure}

\subsection{Error thresholds}
In the absence of fake agents $(\epsilon =0)$ as in \cite{bhw}, $\eta=1$ and so cascading to a $Y (N)$ cascade requires at least two consecutive $Y$'s ($N$'s). However, in the presence of fake agents, $\eta < 1$. In this case, even a single $N$ after a $Y$ could trigger a $N$ cascade. On the other hand, as $\epsilon$ increases and reduces $\eta$, a greater number of consecutive $Y$'s $(\geq 2)$ are required to cause a $Y$ cascade. This is characterized in the following lemma.  
\begin{lemma}
Let $\alpha = p / (1-p) $. For $r \in \mathbb{N}$, define the increasing sequence of thresholds \scalebox{0.95}{$\{ \epsilon_r \}_{r=1}^{\infty}$,} with the $r^{\text{th}}$ threshold $\epsilon_r$ given by
\begin{align}
\epsilon_r = \frac{\alpha - \alpha^{\frac{1}{r}} }{\alpha^{\frac{1}{r}+1} -1}. \label{eps_thesh_eqn}
\end{align}
Define $\mathcal{I}_r \triangleq [ \epsilon_r, \epsilon_{r+1} ) $ as the $r^{\text{th}}$ $\epsilon$–interval. Then for $\epsilon \in \mathcal{I}_r$, at least $r+1$ consecutive $Y$'s are necessary for a $Y$ cascade to begin.
\label{lemma2}
\end{lemma}
\vspace{-3mm}

The proof follows by noting that, if a cascade has not begun immediately after an $N$ is observed, then the rightmost position that the random walk can be in, is in state $0$. Starting from state $0$ therefore gives a lower bound on the number of consecutive $Y$'s required to begin a $Y$ cascade. From here, $r+1$ consecutive $Y$'s would be needed to begin a $Y$ cascade when  \scalebox{0.9}{$ \frac{1}{r+1}< \eta \leq \frac{1}{r}$}.  This inequality implies that $\epsilon \in \mathcal{I}_r = [ \epsilon_r, \epsilon_{r+1} )$ where $\epsilon_r$ is the $r^{\text{th}}$ threshold, defined in \eqref{eps_thesh_eqn}. Here, recall from \eqref{eta} that $ \eta$ is a function of $\epsilon$ and $p$.
\begin{remark}
For $\epsilon \in \mathcal{I}_r$, starting from state $0$, $r+1$ consecutive $Y$'s start a $Y$ cascade. Further, the integer $r$ satisfies $r = \left\lfloor 1 / \eta \right\rfloor $.
\label{remark1}
\end{remark}

\begin{figure}[t]
\centering
\includegraphics[width=0.80\linewidth]{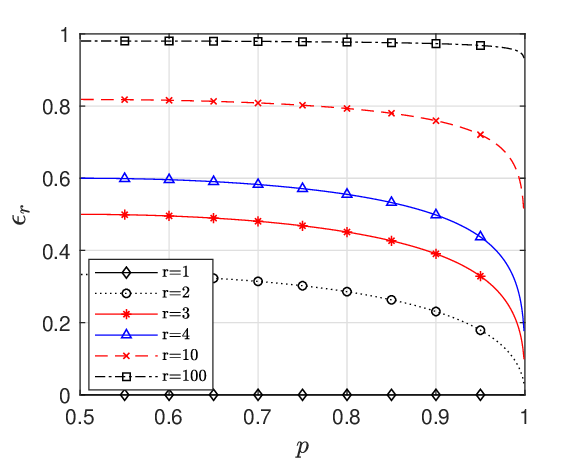} 
\setlength{\abovecaptionskip}{4pt}
\setlength{\belowcaptionskip}{-8pt}
\caption{\small Thresholds $\epsilon_r$ for the indicated values of $r$ versus $p$.}
\label{eps_thresh} 
\end{figure}

Figure \ref{eps_thresh} shows the thresholds $\epsilon_r$ varying with $p$, for different values of $r$. For a fixed $\epsilon$, we see that as the signal quality ($p$) improves, more consecutive $Y$'s are required for a $Y$ cascade to begin. This is because, an increase in $p$ increases the information contained in a $Y$, but not as much as the corresponding increase in the information contained in a $N$. Further, note that as $\epsilon \rightarrow 1$, $r \rightarrow \infty$ which implies that infinitely many consecutive $Y$'s are required for a $Y$ cascade to begin. Equivalently, the information contained in a $Y$ observation becomes negligible. We further investigate learning in this asymptotic scenario in Section \ref{sec:learning_limeps_1}.

\vspace{-1mm} 
\subsection{$\bm{Y}$ cascade probability, $\bm{\mathbb{P}_{Y\text{-cas}}}$} \label{subsec:Ycas_prob} 
In this subsection, we will compute the probability of absorption of $\{h_n\}$ to a $Y$ cascade (into the right absorption region) given $V$, namely $\mathbb{P}_{Y\text{-cas}}^V$. Any probability stated in the remainder of this section is assumed to be conditional under $V$ through its dependence on $p_f$; we thereby drop $V$ from its superscript for notational convenience. Consider the iterative method depicted in Figure \ref{iter_struct} that describes all possible sequences that can lead to a $Y$ cascade. For this process, we assume $\epsilon \in \mathcal{I}_r \smallsetminus \{\epsilon : 1/\eta \in \mathbb{Q}\}$ for some $r = 1,2, \ldots$. We do this to avoid the case of rational values of $1 / \eta$ which could result in certain special sequences (having two consecutive $N$'s) with non-zero probability, that are not enumerated in Figure \ref{iter_struct}.\footnote{These special sequences can be readily incorporated into the enumerations in Figure \ref{iter_struct}, but would disrupt the recursive pattern. For simplicity, we avoid such values of $\epsilon$} For the case of rational-valued $1 / \eta$, recall that the absorption probabilities can be obtained directly as solutions to a system of linear equations.





\vspace{0.5mm}
To begin the iterative method, we initialize Stage $1$ with $r_1 = r+1 $. Now, starting from state $0$, consider the sequences shown in Stage $1$ of Figure \ref{iter_struct}. The first sequence of $r_1$ consecutive $Y$'s, denoted by $Y^{r_1}$, clearly enters the right absorption region (Remark \ref{remark1}), and so $r_1 \eta \in [1,1+\eta]$. The rest of the sequences, each of length $r_1+1$, are simply permutations of each other that contain only a single $N$. This is because two $N$'s or more are not possible without entering the left absorption region. Now, each of these $r_1$ distinct sequences results in the same net right shift of $r_1 \eta -1$, which ends in the region $[0,\eta]$ as we know that $r_1 \eta \in [1,1+\eta]$. This completes Stage $1$. From here, it would take either $r$ or $r+1$ consecutive $Y$'s to enter the right absorption region. Let this value be denoted by $r_2$. The sequences in Stage $2$ can then be enumerated exactly as in the first stage, except that $r_2$ now replaces $r_1$. Now, unless there are $r_2$ consecutive $Y$'s, the sequences of Stage $2$ again end in the region $[0,\eta]$, and then the process continues to the next stage. Here, $r_n$ denotes the number of consecutive $Y$'s required to enter the right absorption region in the $n^{\text{th}}$ stage. In this manner, all sequences that lead to a $Y$ cascade are enumerated. 

\begin{figure}[t]
\centering
\begin{tikzpicture}[scale=2]
\begin{scope}[shift={(-3,0)}]
\draw [dashed, decoration={markings,mark=at position 1 with {\arrow[scale=1,>=stealth]{>}}},postaction={decorate}] (0.1,0) -- (0.3,0) -- (0.3,0.2); \node at (0.4,0.3) {\scalebox{0.8}{$Y$ cascade}};
\draw [rounded corners, decoration={markings,mark=at position 1 with {\arrow[scale=2,>=stealth]{>}}},postaction={decorate}] (0.8,-0.6) -- (1.05,-0.6);
\node at (-0.1,0) {$Y^{r_1}$}; 
\node at (0,-0.2) {$N \; Y^{r_1}$};  \draw [rounded corners] (0.3,-0.2) -- (0.85,-0.2)--(0.85,-0.5);
\node at (0.2,-0.4) {$Y \; N \; Y^{r_1-1}$}; \draw [rounded corners] (0.7,-0.4) -- (0.85,-0.4);
\node at (0.25,-0.6) {$Y^2 \; N \; Y^{r_1-2}$}; 
\draw [dotted] (0,-0.7) -- (0,-1.1);
\node at (0.23,-1.2) {$Y^{r_1-1} \; N \; Y $}; \draw [rounded corners] (0.7,-1.2) -- (0.85,-1.2)--(0.85,-0.5);
\draw [dashed] (-0.35,0.1) -- (-0.35,-1.3);
\draw [dashed] (1.05,0.1) -- (1.05,-1.3);
\node at (0.23,-1.7) {Stage $(1)$};
\end{scope}
\begin{scope}[shift={(-1.6,0)}]
\draw [dashed, decoration={markings,mark=at position 1 with {\arrow[scale=1,>=stealth]{>}}},postaction={decorate}] (0.1,0) -- (0.3,0) -- (0.3,0.2); \node at (0.4,0.3) {\scalebox{0.8}{$Y$ cascade}};
\draw [rounded corners, decoration={markings,mark=at position 1 with {\arrow[scale=2,>=stealth]{>}}},postaction={decorate}] (0.8,-0.6) -- (1.05,-0.6);
\node at (-0.1,0) {$Y^{r_2}$};
\node at (0,-0.2) {$N \; Y^{r_2}$};  \draw [rounded corners] (0.3,-0.2) -- (0.85,-0.2)--(0.85,-0.5);
\node at (0.2,-0.4) {$Y \; N \; Y^{r_2-1}$}; \draw [rounded corners] (0.7,-0.4) -- (0.85,-0.4);
\node at (0.25,-0.6) {$Y^2 \; N \; Y^{r_2-2}$}; 
\draw [dotted] (0,-0.7) -- (0,-1.1);
\node at (0.23,-1.2) {$Y^{r_2-1} \; N \; Y $}; \draw [rounded corners] (0.7,-1.2) -- (0.85,-1.2)--(0.85,-0.5);
\draw [dashed] (1.05,0.1) -- (1.05,-1.3);
\node at (0.23,-1.7) {Stage $(2)$};
\node at (1.3,-0.6) {$\ldots$};
\end{scope}
\end{tikzpicture}
\caption{\small An enumeration of all possible sequences that would lead to a $Y$ cascade. The term $Y^{t}$ represents a sequence of $t$ consecutive $Y$'s. The sequence $\{r_n \}$ is defined as per \eqref{r_updates}.}  
\setlength{\belowcaptionskip}{10pt}
\label{iter_struct}
\vspace{-3.3mm}
\end{figure}





\vspace{0.5mm}
Let $P_n$ denote the probability of entering the right absorption region given that the sequence has not terminated in a $Y$ cascade before the $n^{\text{th}}$ stage. The following recursion then holds.
\begin{align}
P_n = p_f^{r_n} \big[ 1+ r_n (1-p_f) P_{n+1} \big], \;\; \text{for} \; n=1,2,\ldots 
\label{reccur}
\end{align}
and the probability of a $Y$ cascade, denoted by $\mathbb{P}_{Y\text{-cas}}$ is:
\begin{align}
\mathbb{P}_{Y\text{-cas}} (\epsilon) = P_1, \quad \text{for} \;\; \epsilon \in \mathcal{I}_r \smallsetminus \{\epsilon : 1/\eta \in \mathbb{Q}\}.
\label{PYcas}
\end{align}
Here, while $r_1 =r+1$, successive values of $r_i$ for $i=2,3,\ldots$ can be obtained from $r_1$ using the updates:
\begin{align}
r_n = \begin{cases}
r, \quad &\text{if} \;\; \sum_{i=1}^{n-1} (r_i \eta -1) +r \eta > 1, \\
r+1, \quad & \text{o.w.}
\end{cases}
\label{r_updates}
\end{align}

Since \eqref{reccur} is an infinite recursion, to compute  $\mathbb{P}_{Y\text{-cas}}$ in practice, we truncate the process to a finite number of iterations $M$. To this end, we first assume that $P_{M+1}=1$. Next, we use \eqref{reccur} to  successively compute $P_k$ while $k$ counts down from $M$ to $ 1$, performing a total of $M$ iterations. We denote the obtained value as \scalebox{0.95}{$\mathbb{P}_{Y\text{-cas}}^M$}. The following theorem shows that \scalebox{0.95}{$\mathbb{P}_{Y\text{-cas}}^M$} is in fact a tight upper bound to $\mathbb{P}_{Y\text{-cas}}$ as $M \rightarrow \infty$. Moreover, the difference \scalebox{0.9}{$\mathbb{P}_{Y\text{-cas}}^M - \mathbb{P}_{Y\text{-cas}}$} decays to zero at least as fast as $0.5^M$, in the number of iterations $M$. Refer to Appendix \ref{app:Thm1_proof} for a detailed proof.

\begin{theorem}
Let $\epsilon \in \mathcal{I}_r \smallsetminus \{\epsilon : 1/\eta \in \mathbb{Q}\} $ for some $r=1,2,\ldots$, with $p_f$ denoting the probability of a $Y$. Then, for any $M=1,2,\ldots$,
\begin{align*}
0 \, \leq \; \mathbb{P}_{Y\text{-cas}}^M (\epsilon) - \mathbb{P}_{Y\text{-cas}} (\epsilon) \; \leq  \, k^M,
\end{align*}
where $k \triangleq (r+1)(1-p_f) p_f^r$. Further, for any $p \in (0.5,1)$ and $\epsilon \in [0,1)$, $k$ satisfies $0 < k \leq 1/2$.
\label{Thm:1}
\end{theorem}





\begin{figure}[h]
\centering
\includegraphics[trim = 0 0 0 2.5mm, width=\linewidth, clip]{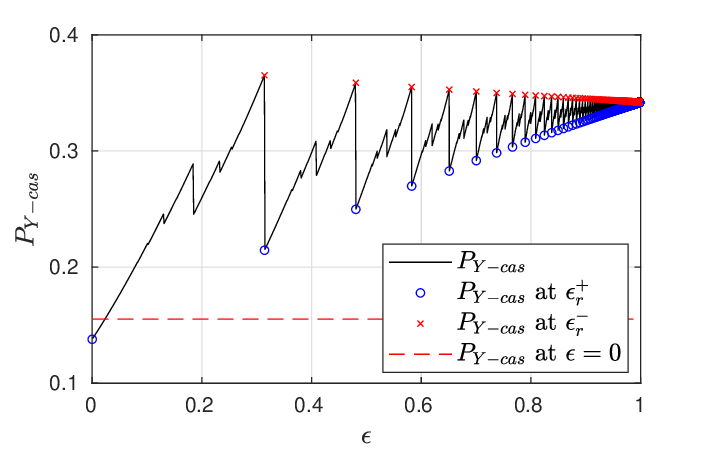}
\setlength{\abovecaptionskip}{-5pt}
\setlength{\belowcaptionskip}{-3pt}
\caption{\small Probability of $Y$ cascade as a function of $\epsilon$ for $V=B$ and $p=0.7$.}
\label{PYcas_fig}
\end{figure}

Figure \ref{PYcas_fig} shows a plot of $\mathbb{P}_{Y\text{-cas}}$ with respect to $\epsilon$, for the case $V=B$ with $p=0.7$. The plot uses $M=10$ which gives an error of less than $10^{-3}$. It can be seen that in the $r^{\text{th}}$ $\epsilon$–interval $\mathcal{I}_r$, $\mathbb{P}_{Y\text{-cas}}$ increases with $\epsilon$, but with infinitely many discontinuities (where $\mathbb{P}_{Y\text{-cas}} (\epsilon)$ decreases). This is in contrast with the unbiased noise model in \cite{Tho}, where it increases continuously over a similar interval. This distinction is a consequence of the state-space of $\{h_n\}$ being infinite, unlike in \cite{Tho}. Despite the discontinuities, $\mathbb{P}_{Y\text{-cas}}$ achieves the minimum (marked by $\circ$) and maximum (marked by $\times$) values at the edge points of $\mathcal{I}_r$, \emph{i.e.}, $\epsilon_{r}^{+}$ and $\epsilon_{r+1}^{-}$, respectively. Further, note that the relatively larger drops in $\mathbb{P}_{Y\text{-cas}} $ observed in Figure \ref{PYcas_fig} occur exactly at the threshold points $\{ \epsilon_r \}_{r=1}^{\infty}$. Here, counter to expectation, a slight increase in $\epsilon$ beyond $\epsilon_r$ causes a significant decrease in the probability of a $Y$ cascade. The same characteristic is exhibited when $V=G$, which is shown in Figure \ref{PYcas_VG_fig} in Appendix \ref{app:PYcas_VGood}. In Section \ref{subsec:abrupt_drop_Pycas}, we show that closed-form expressions for $\mathbb{P}_{Y\text{-cas}}$ as $\epsilon \rightarrow \epsilon_r^+$ and $\epsilon \rightarrow \epsilon_r^-$ can be obtained (refer to \eqref{eps_plus} and \eqref{eps_minus}); thereby quantifying its drop at each threshold $\epsilon_r$. 


\vspace{0.3mm}
In Figure \ref{iter_struct_Ncas} of Appendix \ref{app:PNcas}, we outline an iterative method similar to that in Figure \ref{iter_struct}, to compute the probability of absorption of $\{h_n\}$ to a $N$ cascade, namely $\mathbb{P}_{N\text{-cas}}$ with arbitrary precision. Having already computed $\mathbb{P}_{Y\text{-cas}}$, this alternate process may seem redundant due to the fact that $\mathbb{P}_{N\text{-cas}} = 1-\mathbb{P}_{Y\text{-cas}}$. However, later in Subsection \ref{subsec:finite_agent}, we will be using both the iterative methods depicted by Figures \ref{iter_struct} and \ref{iter_struct_Ncas} to compute exact probabilities of a $Y$ and $N$ cascade respectively, that may begin at or before a finite agent index $n$. We denote the probability that a $Y$ $(N)$ cascade occurs by the $n^{\text{th}}$ agent index by $v_n^{*V}$ $(u_n^{*V})$. Note that agent $n+1$ always has a non-zero probability $1-v_n^{*V}-u_n^{*V}$ of following its private signal $S_{n+1}$, hence $u_n^{*V} \neq 1 - v_n^{*V}$. As a result, $u_n^{*V}$ cannot be inferred by knowing $v_n^{*V}$, and vice-versa. So, both the iterative methods are required to obtain cascade probabilities for a finite $n$.


\vspace{-1.5mm}
\subsection{Quantifying the abrupt changes in $\bm{\mathbb{P}_{Y\text{-cas}}}$ at error thresholds} \label{subsec:abrupt_drop_Pycas}
We now quantify $\mathbb{P}_{Y\text{-cas}}$ as $\epsilon$ tends to each threshold point $ \epsilon_r $. For this, we state the following lemma, with the proof provided in Appendix \ref{app:lemma_ri_proof}.
\begin{lemma}
For all $i \geq 2$, $r_i$ in \eqref{r_updates} satisfies: $r_i \rightarrow r$ as $\epsilon \rightarrow \epsilon_r$.
\label{lemma_for_ri}
\end{lemma}

It follows from Lemma \ref{lemma_for_ri} that as $\epsilon \rightarrow \epsilon_r$, the recursion in \eqref{reccur} results in the same infinite computation to obtain $P_i$ as for $P_{i+1}$, for all $i \geq 2$. Thus, all $P_i$ for $i \geq 2$ have the same value which satisfies: \scalebox{0.92}{$P_i = p_f^{r} \big[ 1+ r (1-p_f) P_{i} \big]$}. Solving this equation for $i=2$ gives the value for $P_2$ which when used in equation \eqref{reccur} for $n=1$ yields $P_1$, \emph{i.e.}, $\mathbb{P}_{Y\text{-cas}}$. However, note that while solving equation \eqref{reccur}, $r_1=r+1$ for $\epsilon=\epsilon_r^+$ whereas $r_1=r$ for $\epsilon=\epsilon_r^{-}$. This corresponds to the following two different values of $\mathbb{P}_{Y\text{-cas}}$ as $\epsilon \rightarrow \epsilon_r$:
\begin{align}
\mathbb{P}_{Y\text{-cas}} (\epsilon_r^{+}) &= \text{\scalebox{0.9}{$\displaystyle p_f^{r+1} \frac{ 1+ (1-p_f)  p_f^{r} }{1- r(1-p_f) p_f^{r}}$}},
\label{eps_plus} \\
\mathbb{P}_{Y\text{-cas}} (\epsilon_r^{-}) &= \text{\scalebox{0.9}{$\displaystyle p_f^{r} \frac{ 1}{1- r(1-p_f) p_f^{r}}$}}.  \label{eps_minus}
\end{align}
Hence, the fractional decrease in $\mathbb{P}_{Y\text{-cas}}$ that occurs abruptly at $\epsilon_r$, defined as \scalebox{0.9}{$\delta_r = \big[ \mathbb{P}_{Y\text{-cas}} (\epsilon_r^{-}) - \mathbb{P}_{Y\text{-cas}} (\epsilon_r^{+})  \big] / \mathbb{P}_{Y\text{-cas}} (\epsilon_r^{-})  $} is
\begin{align}
\delta_r = (1-p_f)\, (1-p_f^{r+1}).
\label{delta_r}
\end{align}
\begin{property}  \label{prop3}
If the possibility of fake agents equals the $r^{th}$ $\epsilon$-threshold, $r=2,3, \ldots$, then for any $V,$ a further marginal increase in fake agents reduces the chances of their preferred \scalebox{1}{$(Y)$} cascade by a factor of \scalebox{1}{$\delta_r$}, rather than increasing it.
\end{property}


The above property implies that, increasing $\epsilon$ just above $\epsilon_r$ is in fact counter-productive for the fake agents. The intuition underlying Property \ref{prop3} is as follows. Recall from Lemma 2 that an increase in $\epsilon$ from $\epsilon_r^{-}$ to $\epsilon_r^{+}$ increases the least number of consecutive $Y$'s required to begin a $Y$ cascade by one. This implies that any observation sequence that ends in a $Y$ cascade for $\epsilon = \epsilon_r^{+}$ gaurantees a $Y$ cascade for $\epsilon = \epsilon_r^{-}$. However, the opposite is not true. Therefore, increasing the noise value from $\epsilon_r^{-}$ to $\epsilon_r^{+}$ results in a drop in the $Y$ cascade probability. 

Lastly, it can be verified from \eqref{delta_r} that as $r \rightarrow \infty$, $\delta_r \rightarrow 0$. This is depicted in Figure \ref{PYcas_fig} where the sequences \scalebox{0.92}{$\{ \mathbb{P}_{Y\text{-cas}} (\epsilon_r^{-}) \}$} and \scalebox{0.92}{$\{ \mathbb{P}_{Y\text{-cas}} (\epsilon_r^{+}) \}$}, marked by $\times$ and $\circ$ respectively, converge to a limiting value as $r \rightarrow \infty$. Note that $r \rightarrow \infty$ implies the limiting scenario of $\epsilon \rightarrow 1$. We investigate this scenario further in Section \ref{sec:learning_limeps_1}.

\vspace{-0.5mm}
\subsection{Effect of fake agents on $\bm{\mathbb{P}_{Y\text{-cas}}}$ at low values of $\bm{\epsilon$}}
\label{subsec:PYcas_low_epsilon}
\begin{figure}[h]
\centering
\includegraphics[trim = 0 0 0 2.5mm, width=\linewidth, clip]{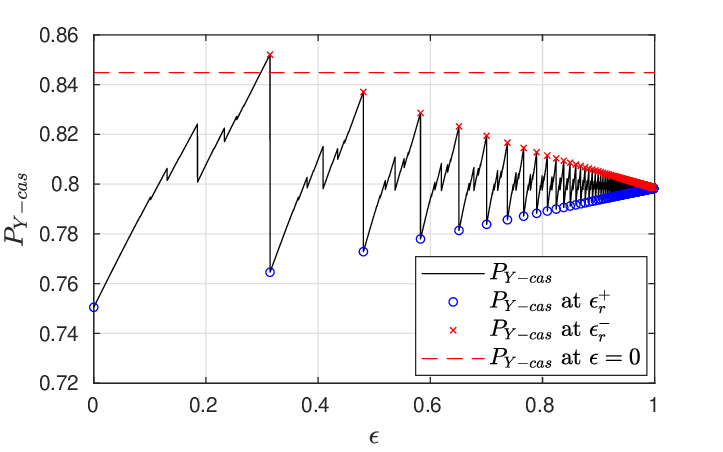}
\setlength{\abovecaptionskip}{-5pt}
\setlength{\belowcaptionskip}{-8pt}
\caption{\small Probability of $Y$ cascade as a function of $\epsilon$ for $V=G$ and $p=0.7$.}
\label{PYcas_VG_fig}
\end{figure}

Next, we consider the cascade behaviour for low values of $\epsilon$. In the absence of fake agents $(\epsilon=0)$ as in \cite{bhw}, $\eta =1$. It then follows from Figure \ref{random_walk} that $\{h_n\}$ has a finite state-space $\{-2,-1,0,1,2\}$, and a $Y$ $(N)$ cascade starts when $h_n = 2$ $(-2)$. Here, solving for  $\mathbb{P}_{Y\text{-cas}}$ gives
\begin{align}
\hspace{-1mm} \mathbb{P}_{Y\text{-cas}} (0) = \begin{cases} p^2 / \big[ p^2 + (1-p)^2 \big],  &\text{for} \; V=G, \\
(1-p)^2 / \big[ p^2 + (1-p)^2 \big], \; &\text{for} \; V=B.
\end{cases} 
\label{PYcas_zero}
\end{align}
In Figure \ref{PYcas_fig} and Figure \ref{PYcas_VG_fig} (Appendix \ref{app:PYcas_VGood}), this is indicated as a baseline to compare it with $\mathbb{P}_{Y\text{-cas}}$ for any $\epsilon \in (0,1)$, when $V=B$ and $V=G$, respectively. Observe that for low values of $\epsilon$, $\mathbb{P}_{Y\text{-cas}}$ is smaller than when fake agents are absent. The next theorem formalizes this property for any $p$ and any $V.$
\begin{theorem}
Given the private signal quality $p \in (0.5,1)$, and the item's true value $V \in \{G,B\}$, there exists some $\underline{\epsilon} = f(V,p) > 0$ such that
\begin{align}
\mathbb{P}_{Y\text{-cas}} (\epsilon) < \mathbb{P}_{Y\text{-cas}} (0), \quad \forall \;\; \epsilon \in (0,\underline{\epsilon}).
\label{equation_eps_thm}
\end{align}
\label{eps_thm}
\end{theorem}
\vspace{-5mm}
\noindent From the point-of-view of the fake agents, the above theorem implies the following property.
\begin{property}  \label{prop4}
If fake agents occur with a probability of less than $\underline{\epsilon}$, then the effect that their presence has on the honest buyers is opposite to what they had intended. That is, they reduce the chances of their preferred $(Y)$ cascade instead of increasing it.
\end{property}


\noindent Likewise, Theorem \ref{eps_thm} implies that if $V=B$, then the honest buyers benefit from the presence of fake agents when $\epsilon < \underline{\epsilon}$. Otherwise, if $V=G$, then they are harmed by such a presence of fake agents.



\begin{IEEEproof} Note that as $\epsilon \rightarrow 0$, the limiting value of $\mathbb{P}_{Y\text{-cas}}$ can be obtained from \eqref{eps_plus} with $r=1$ and $p_f \rightarrow 1-p$ for $V=B$, whereas $p_f \rightarrow p$ for $V=G$. This gives
\begin{align}
\hspace{-1mm} \text{\scalebox{0.95}{$\lim_{\epsilon \rightarrow 0 } \mathbb{P}_{Y\text{-cas}} (\epsilon) =$}} \begin{cases}
\text{\scalebox{0.9}{$\displaystyle (1-p)^2 \frac{1+p(1-p)}{1-p(1-p)} $}}, \quad &\text{for} \;\; V=B, \vspace{2mm}\\
\text{\scalebox{0.9}{$\displaystyle p^2 \frac{1+p(1-p)}{1-p(1-p)} $}}, \quad &\text{for} \;\; V=G.
\end{cases} 
\label{lim_eps_to_0}
\end{align}
In Figure \ref{PYcas_fig}, this limiting value is marked by $\circ$ at $\epsilon =0$. By comparing this expression with the one in \eqref{PYcas_zero} for the corresponding values of $V,$ we have
\begin{align*}
\lim_{\epsilon \rightarrow 0 } \mathbb{P}_{Y\text{-cas}} (\epsilon) < \mathbb{P}_{Y\text{-cas}} (0), \quad \forall \; p \in (0.5,1), V \in \{G,B\}.
\end{align*}
Thus, there exists some $\underline{\epsilon} > 0$ such that \eqref{equation_eps_thm} holds true. 
\end{IEEEproof}

From the above proof, it follows that Theorem \ref{eps_thm} is a consequence of the discontinuity in $\mathbb{P}_{Y\text{-cas}}$ at $\epsilon = 0$, for any given $V$. This contrasts with the unbiased noise model in \cite{Tho}, where there is no discontinuity in the cascade probabilities at $\epsilon = 0$ (no noise). Once again, this distinction results from the state-space of $\{h_n\}$ being infinite, unlike in \cite{Tho}.


\vspace{-3mm}
\subsection{Probability of cascades for finite agent arrivals}  \label{subsec:finite_agent}
In this subsection, we use the iterative methods depicted in Figure \ref{iter_struct} and Figure \ref{iter_struct_Ncas} (in Appendix \ref{app:PNcas}) to compute exact probabilities for correct and wrong cascades to begin within the arrival of a finite number of agents. For the random walk $\{h_n\}$ in Figure \ref{random_walk}, given the true value $V \in \{G,B\}$, let $v_n^{*V}$ $(u_n^{*V})$ be the probability of being absorbed by the right (left) absorption region by the $n^{\text{th}}$ time-step. As the time-steps of $\{h_n\}$ correspond to the agents' indices, $v_n^{*V}$  and $u_n^{*V}$ are respectively the probabilities that a correct and wrong cascade occurs by the $n^{\text{th}}$ arrival, under $V=G$. Vice-versa holds when $V=B$. Now, let $v_n^V$ $(u_n^V)$ be the probability that $\{h_n\}$ enters the right (left) absorption region exactly at the $n^{\text{th}}$ time-step; it follows that
\begin{align}
v_n^{*V} = \sum_{i=1}^{n} v_i^V  \quad \text{and} \quad u_n^{*V} = \sum_{i=1}^{n} u_i^V.
\label{v*n_u_*n}
\end{align}
Figure \ref{PYcas_fig} shows that all sequences that enter the right absorption region (\emph{i.e.} end in a $Y$ cascade) in Stage $(j)$ terminate with a $Y^{r_j}$ and have the same length: $l_j \triangleq  r_j +\sum_{i=1}^{j-1} (r_i+1) $, for $j = 1,2,\ldots$. This yields the values for $\{ v_n \}$ as follows:
\begin{align}
v_n = \begin{cases}  \text{\scalebox{0.9}{$p_f^{r_j} \displaystyle \prod_{i=1}^{j-1} r_i (1-p_f) p_f^{r_i},$}} \quad &\text{if} \; n=l_j, \\
0, \quad &\text{o.w.} 
\end{cases}
\label{v_n}
\end{align}
In a similar manner, by observing Figure \ref{iter_struct_Ncas}, we see that among all sequences that end in a $N$ cascade in Stage $(j)$, the $t$ sequences that terminate in any allowable permutation of $ N Y^{t-1} N$  have the same length: $  t+1 +\sum_{i=1}^{j-1} (r_i+1) = l_{j-1}+t+2 $, for $t=1,2,\ldots,r_j$ and $j = 1,2,\ldots$. This yields the values for $\{ u_n \}$ as follows: 
\begin{align}
\mspace{-10mu} \text{\scalebox{0.9}{$u_n =$}} \!  \begin{cases} \text{\scalebox{0.9}{$\displaystyle t p_f^{t-1} (1-p_f)^2 \prod_{i=1}^{j-1} r_i (1-p_f) p_f^{r_i},$}} \! \! \!  &\text{if} \;  \text{\scalebox{0.88}{$n=l_{j-1} \!+t+2,$}} \\
0,  &\text{o.w.} 
\end{cases}
\label{u_n}
\end{align}

Now, as $n \rightarrow \infty$, the asymptotic value of the quantity $ v_{n-1}^{*V} $ $( u_{n-1}^{*V})$ refers to the probability that a $Y$ $(N)$ cascade occurs eventually. Thus,
\begin{align}
\hspace{-1mm} \text{\scalebox{0.95}{$\underset{n \rightarrow \infty}{\lim} v_n^{*V} = \mathbb{P}^V_{Y\text{-cas}}$}}  \; \; \text{and} \; \; \text{\scalebox{0.95}{$\underset{n \rightarrow \infty}{\lim} u_n^{*V} = \mathbb{P}^V_{N\text{-cas}} \overset{(a)}{=} 1- \mathbb{P}^V_{Y\text{-cas}}$}},
\label{lim_vn_un}
\end{align}
where equality $(a)$ holds since it can be shown that $\{h_n\}$ is absorbed into either of the cascades w.p. $1$.

\section{Welfare for ordinary agents}
\label{sec:Exp_pay-off}
In this section, we analyze the expected pay-off or welfare of agent $n$, $\pi_n$ if it is ordinary (rational), i.e., if it takes a pay-off optimal action. Recall from Section \ref{sec:model} that $\pi_n=0$ if \scalebox{1}{$A_n = N$} whereas $\pi_n= 1/2$ or $-1/2$ if \scalebox{1}{$A_n=Y$} depending on whether \scalebox{0.96}{$V=G$ or $B$,} respectively. Now, Figure \ref{random_walk} shows that it takes at least two steps to begin a cascade. This implies that the first two agents always follow their private signals, and hence have the same welfare given by
\begin{align}
\hspace{-2mm} \mathbb{E}[\pi_n] &= \mathbb{P} (A_n = Y \vert V=G) \frac{1}{4} - \mathbb{P} (A_n = Y \vert V=B) \frac{1}{4}, \label{Exp_pi_n} \\
&= (2p-1)/4 \triangleq F, \quad \text{for} \;\; n \in \{1,2\}.
\label{F}
\end{align}
In fact, $F$ defined in \eqref{F} refers to the welfare for any agent $n$, if $A_n$ always follows $S_n$ disregarding the optimal decision rule in \eqref{bayes_decison}, i.e., \scalebox{0.97}{$\mathbb{E}[\pi_n \vert A_n \; \text{always follows} \; S_n] = F,$} for all $n$. However, for agents $n > 2$, the unconditional welfare must also account for the possibilities that the observed history $\mathcal{H}_{n-1}$ could cause agent $n$ to cascade to a $Y$ or a $N$, which can be expressed as follows:
\begin{flalign}
\text{ \scalebox{0.92}{$\mathbb{E}[\pi_n] = F + \frac{1}{4} \left[ (1-p) v_{n-1}^{*G} - p v_{n-1}^{*B} + (1-p) u_{n-1}^{*B} -p u_{n-1}^{*G} \right] $}}. &&
\label{E_pi_n}
\end{flalign}
The above equation explicitly relates agent $n$'s welfare to the probabilities of $Y$ and $N$ cascades resulting from its history $\mathcal{H}_{n-1}$ under both $V=G$ and $B$.

\vspace{0.2mm}
Now, assuming that agent $n$ can choose to observe only a certain subset of the available observations $I_n = \{O_1,\ldots ,O_{n-1}, S_n\}$, the next theorem shows that its welfare cannot get worse if more elements are added to this subset. This implies the non-redundancy of the observations in $I_n$ as agent $n$ can never achieve a welfare higher than in \eqref{E_pi_n} by ignoring any of the observations in $I_n$.

\begin{theorem}
Let $\mathcal{J}_n$ be the collection of all subsets of the observations $\{O_1,\ldots ,O_{n-1}, S_n\}$ that are available to agent $n$. Let $\mathbb{E}[\pi_n \vert J]$ refer to the optimal welfare achieved by agent $n$ by only observing the set $J \in \mathcal{J}_n$. Then, for any two sets $J,K \in \mathcal{J}_n$ such that $K \subset J$, we have $\mathbb{E}[\pi_n \vert J] \geq \mathbb{E}[\pi_n \vert K]$.
\label{Blackwell}
\end{theorem}
Thereom \ref{Blackwell} can be proved by applying the celebrated Blackwell's Theorem \cite{leshno1992elementary} which implies that it is sufficient to show that the signals from observing the smaller set $K$ are obtained as a stochastic mapping (garbling) of the signals from the larger set $J$. Let $\bar{J}$ and $\bar{K}$ be the $j$ and $k$-length random vectors $(j > k)$ corresponding to the observations sets $J$ and $K$, respectively, such that the two vectors share the first $k$ elements. Then, the desired mapping is given by $\bar{K} = G \bar{J}$, where $G$ is a $k \times j$ diagonal matrix. Then, Blackwell's result for the corresponding optimal welfares states that $\mathbb{E}[\pi_n \vert J] \geq \mathbb{E}[\pi_n \vert K]$. 

As a corollary of Theorem \ref{Blackwell}, the next property shows that the welfare under complete observation, i.e., $\mathbb{E}[\pi_n]$ is monotonic in the agents' indices.

\begin{property}
The welfare of each agent is at least equal or greater than the welfare of its predecessors. Thus, $E[\pi_n] \geq F$ and is non-decreasing in $n$.
\label{prop:monotonic}
\end{property}
To see this property, consider two consecutive agents, $n-1$ and $n$. Under the informational equivalance of  their private signals: $S_{n-1}$ and $S_n$, we have $I_{n-1},I_n \in \mathcal{J}_n$ and $I_{n-1} \subset I_{n}$. By applying Theorem \ref{Blackwell}, this implies that $\mathbb{E}[\pi_n] \geq \mathbb{E}[\pi_{n-1}]$; and  $\mathbb{E}[\pi_n] \geq F$ for all $n$ follows from \eqref{F}.

\begin{figure}
\centering
\includegraphics[trim = 0cm 0mm 0cm 2.5mm, width=\linewidth, clip]{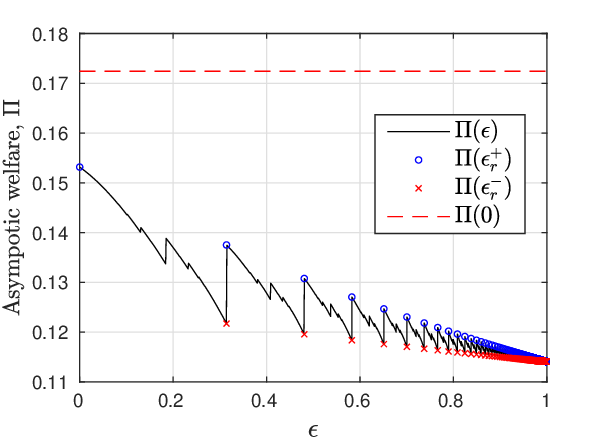}
\setlength{\abovecaptionskip}{-5pt}
\setlength{\belowcaptionskip}{-10pt}
\caption{\small Asymptotic welfare as a function of $\epsilon$ for $p=0.7$.}
\label{Exp-pay-off_fig}
\end{figure}

\vspace{0.3mm}
Next, by taking the limit as $n \rightarrow \infty$ of \eqref{E_pi_n} and using \eqref{lim_vn_un}, the asymptotic welfare $\underset{n \rightarrow \infty}{\lim} \mathbb{E} [\pi_n (\epsilon)]$ denoted by $\Pi (\epsilon)$ is given by
\begin{align}
\Pi(\epsilon) &=  \text{\scalebox{0.93}{$\frac{1}{4} \left[ \mathbb{P}_{Y\text{-cas}}^G (\epsilon) - \mathbb{P}_{Y\text{-cas}}^B (\epsilon) \right] \overset{(a)}{=} \frac{1}{4} \big( 1-2 \mathbb{P}_{\text{wrong-cas}} (\epsilon) \big)$}}.
\label{Pi_inf}
\end{align} 
In Step $(a)$, $\mathbb{P}_{\text{wrong-cas}} := \left[ \mathbb{P}_{N\text{-cas}}^G + \mathbb{P}_{Y\text{-cas}}^B \right] /2$ refers to the unconditional probability of a wrong cascade, i.e., the probability that a $N$ cascade occurs and $V=G$ or a $Y$ cascade occurs and $V=B$. Equation \eqref{Pi_inf} implies that an \emph{improved learning}, i.e., a lower probability of wrong cascade results in a higher asymptotic welfare. The probability $\mathbb{P}_{Y\text{-cas}}^V (\epsilon)$ for $V \in \{G,B\}$ can be computed using the recursive method described in Section \ref{sec:markov}, which is outlined by equations \eqref{reccur}, \eqref{PYcas} and \eqref{r_updates}. Then, substituting these obtained values in \eqref{Pi_inf} yields the value for $\Pi(\epsilon)$. Figure \ref{Exp-pay-off_fig} shows a plot of $\Pi(\epsilon)$ with respect to $\epsilon \in (0,1)$, for $p = 0.7$ and compares it with the constant level of $\Pi(0)$ which refers to the asymptotic welfare in the absence of fake agents. Substituting \eqref{PYcas_zero} in \eqref{Pi_expression} gives
\begin{align}
\Pi(0) = (1/4)(2p-1)/ \big[ \mspace{2mu}  p^2+(1-p)^2  \mspace{2mu} \big]. \label{Pi_zero}
\end{align}
It can be observed in Figure \ref{Exp-pay-off_fig} that for $p=0.7$, $\Pi(\epsilon) < \Pi(0)$ for all $\epsilon \in (0,1)$. Further, note that the relatively larger jumps in $\Pi(\epsilon)$ observed in Figure \ref{Exp-pay-off_fig} (marked by $\times$ and $\circ$) occur exactly at the threshold points $\{ \epsilon_r \}_{r=2}^{\infty}$. Here, counter to expectation, a slight increase in $\epsilon$ beyond $\epsilon_r$ causes an abrupt and significant increase in the asymptotic welfare. This abrupt increase does not simply follow from Property \ref{prop3} because a drop in the $Y$ cascade probability improves learning when $V=B$, whereas it worsens learning when $V=G$. Thus, when averaged across $V$, Property \ref{prop3} is not sufficient to imply that learning improves in turn causing the asymptotic welfare to increase at each $\epsilon_r$-threshold. This is unlike the unbiased noise model in \cite{Tho}, where learning at noise thresholds improves for both $V$'s, thereby implying an increase in asymptotic welfare. We propose the following theorem, which shows that such an abrupt increase in welfare at $\epsilon_r $ occurs not only as $n \rightarrow \infty$ but also occurs for every agent $n$. Refer to Appendix \ref{app:welfare_@eps_r} for a detailed proof.

\vspace{-1mm}
\begin{theorem}
Given a fixed private signal quality $p$, for every agent $ n \in \{1,2, \ldots \}$, at any $r^{\text{th}}$ $\epsilon$-threshold $\epsilon_r$, $r \geq 2$,
\begin{align}
\mathbb{E} [ \pi_n(\epsilon_r^+)] - \mathbb{E} [ \pi_n(\epsilon_r^-)] > 0.
\end{align}
\label{thm:welfare_agent_n_@eps_r}
\end{theorem}
\vspace{-5mm}
It follows from Theorem \ref{thm:welfare_agent_n_@eps_r} that for any $p$, there occurs an abrupt increase of $\Delta_r$ in the asymptotic welfare at each of the threshold points $\{ \epsilon_r \}_{r=2}^{\infty}$, where
\begin{align}
\Delta_r := \Pi(\epsilon_r^+) - \Pi(\epsilon_r^-) > 0.  \label{Delta_r_more_than_zero}
\end{align}

Through the relation in \eqref{Pi_inf}, this improvement of $\Delta_r$ in the asymptotic welfare corresponds to an abrupt reduction of $(4 \Delta_r - 1)/2 $ in the wrong cascade probability, $\mathbb{P}_{\text{wrong-cas}}$. The expression for $\Delta_r$ can be obtained by first using equations \eqref{eps_plus} and \eqref{eps_minus} to compute the values: $\mathbb{P}_{Y\text{-cas}}^V (\epsilon)$ for $V \in \{G,B\}$ and $ \epsilon = \epsilon_r^+ $ and $\epsilon_r^-$. Next, substituting these values in \eqref{Pi_inf} to obtain $\Pi(\epsilon_r^+) $ and $ \Pi(\epsilon_r^-)$ and then using \eqref{delta_r} yields
\begin{align}
\Delta_r = \frac{1}{4} \left[ \delta_r^B \mathbb{P}_{Y\text{-cas}}^B (\epsilon_r^-) - \delta_r^G \mathbb{P}_{Y\text{-cas}}^G (\epsilon_r^-) \right].
\label{Delta_r}
\end{align}

\begin{property}  \label{prop6}
If the possibility of fake agents in the history equals the $r^{th}$ $\epsilon$-threshold, $r=2,3, \ldots$, then a further marginal increase in fake agents improves the welfare at every agent index $n \in \{ 1,2,\ldots\}$. Moreover, the asymptotic welfare of the agents improves by $\Delta_r$.
\end{property}


Therefore, increasing $\epsilon$ over the $r^{\text{th}}$ $\epsilon$-threshold is not only counter-productive for the fake agents (due to Property \ref{prop3}), but it also leads to a higher social welfare for every ordinary agent. The intuition underlying Property \ref{prop6} is that when $\epsilon$ increases from $\epsilon_r^{-}$ to $\epsilon_r^{+},$ the drop in the $Y$ cascade probability for $V=B$ (better learning) is more pronounced than the corresponding drop for $V=G$ (worse learning). This is because this drop is a decreasing function of $p_f$, and that $p_f$ for $V=G$ is greater than $p_f$ for $V=B$ (i.e., $a > 1-b$). Therefore,  this increase in $\epsilon$ implies a net decrease in the wrong cascade probability when averaged over $V,$ which as per \eqref{Pi_inf} leads to a higher welfare for rational agents.



\vspace{-2mm}
\section{Learning in the limit} \label{sec:learning_limeps_1}
\begin{figure}[h!]
\vspace{-2mm}
\centering
\subfloat[]{\includegraphics[trim = 0 0 4mm 2.5mm, width=0.5\linewidth, clip]{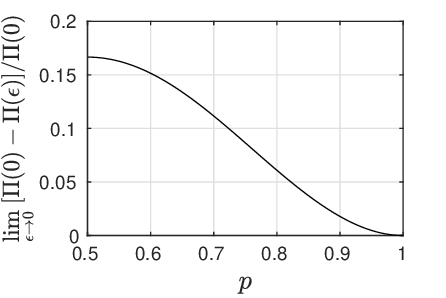}
\label{fig:welfare_decrease_limeps0}}
\subfloat[]{\includegraphics[trim = 0 0 4mm 2.5mm, width=0.5\linewidth, clip]{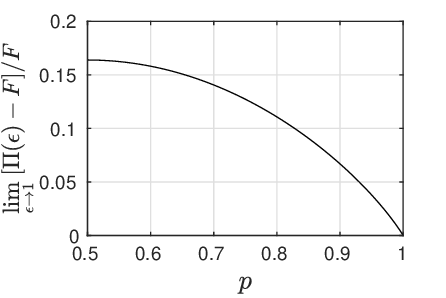}
\label{fig:welfare_increase_limeps1}}
\setlength{\abovecaptionskip}{4pt}
\setlength{\belowcaptionskip}{-3pt}
\caption{\small $(a)$ Fractional reduction in welfare relative to $\Pi(0)$ for varying $p$, as $\epsilon \rightarrow 0$. $(b)$ Fractional increase in welfare relative to $F$ for varying $p$, as $\epsilon \rightarrow 1$.}
\end{figure}

In this section, we evaluate the asymptotic welfare of agents under two limiting regimes of $\epsilon$, namely, $\epsilon \rightarrow 0$ and $\epsilon \rightarrow 1$. In the first case, taking the limit $\epsilon \rightarrow 0$ in \eqref{Pi_expression} and then substituting the expressions obtained in \eqref{lim_eps_to_0} yields  
\begin{align}
\text{\scalebox{0.95}{$\lim_{\epsilon \rightarrow 0 } \Pi = $}} \;  \text{\scalebox{0.9}{$\displaystyle \frac{1}{4} (2p-1)  \frac{1+p(1-p)}{1-p(1-p)} $}}.
\label{Pi_eps_lim_zero}
\end{align} 
Comparing \eqref{Pi_eps_lim_zero} with the asymptotic welfare at $\epsilon=0$ given in \eqref{Pi_zero}, it can be shown that $\lim_{\epsilon \rightarrow 0 } \Pi(\epsilon) < \Pi(0) $ for any $p$. This implies that even an infinitesimal presence of fake agents causes an abrupt deterioration in welfare. Figure \ref{Exp-pay-off_fig} shows this drop in welfare at $\epsilon=0$, where the value $\Pi(0)$ drops to the value $\lim_{\epsilon \rightarrow 0 } \Pi(\epsilon)$, which is marked by $\circ$. Next, we define the fractional reduction (f.r.) in welfare relative to $\Pi(0)$, as $ \displaystyle \lim_{\epsilon \rightarrow 0} \; [\Pi(0)-\Pi(\epsilon)]/\Pi(0)$. We then plot it against $p$ in Figure \ref{fig:welfare_decrease_limeps0} in order to understand the effects of varying the signal quality $p$. It can be shown analytically from the expressions in \eqref{Pi_zero} and \eqref{Pi_eps_lim_zero} that the f.r. in welfare is monotonically decreasing with $p$. The greatest f.r. thus occurs as $p \rightarrow 0$ and is found to equal $1/6$, whereas the f.r. is $0$ in the limit $p \rightarrow 1$.
\vspace{-2mm}


In the case of $\epsilon \rightarrow 1$, the information contained in a $Y$ observation becomes negligible. As a result, an agent would need to observe infinitely many consecutive $Y$'s in his history for him to be convinced of starting a $Y$ cascade. Hence, one would expect that if $V=G$, learning would never occur, whereas if $V=B$, then learning would always occur. However, recall that as $\epsilon \rightarrow 1$, the occurrence of $Y$'s becomes increasingly frequent, \emph{i.e.} $p_f \rightarrow 1$; for both $V=B$ and $G$. This motivates studying $\mathbb{P}_{Y\text{-cas}}$ in this limiting scenario. First, recall that in the process of enumerating all sequences leading to a $Y$ cascade, for $\epsilon \in \mathcal{I}_r$, in each stage $i \geq 2$, $r_i$ is either $r$ or $r+1$. However, $\epsilon \rightarrow 1$ implies $r\rightarrow \infty$, in which case $r \approx r+1$. As a result, the expressions obtained in \eqref{eps_plus} and \eqref{eps_minus} yield the same limiting value as $r\rightarrow \infty$, which also equals $\mathbb{P}_{Y\text{-cas}}$ as $\epsilon \rightarrow 1$. In particular, 
\begin{align}
\lim_{\epsilon \rightarrow 1 } \mathbb{P}_{Y\text{-cas}} (\epsilon) = \lim_{r \rightarrow \infty } \mathbb{P}_{Y\text{-cas}} (\epsilon_r^{+}) \overset{(b)}{=} \lim_{r \rightarrow \infty } \mathbb{P}_{Y\text{-cas}} (\epsilon_r^{-}),   \label{eqn:eps_lim_1}
\end{align}
where Step $(b)$ can also be proved by recalling that $\delta_r \rightarrow 0$ as $r \rightarrow \infty$. By using \eqref{eps_minus} in \eqref{eqn:eps_lim_1}, the limiting probability of a $Y$ cascade, in terms of $\alpha = p / (1-p)$ can  be obtained as:
\begin{align}
\lim_{\epsilon \rightarrow 1 } \mathbb{P}_{Y\text{-cas}} (\epsilon) &= \lim_{r \rightarrow \infty } \text{\scalebox{0.9}{$\displaystyle p_f^{r} \frac{ 1}{1- r(1-p_f) \mspace{1.5mu} p_f^{r}} = \frac{1}{e^t - t}$}};
\label{eps_lim_expression}
\end{align}
where \scalebox{0.95}{$ t = \frac{1}{\alpha-1} \log \alpha$} for \scalebox{0.95}{$V=G$}, and \scalebox{0.95}{$ t= \frac{\alpha}{\alpha-1} \log \alpha$} for \scalebox{0.95}{$V=B$}. A detailed proof of \eqref{eps_lim_expression} is provided in Appendix \ref{app:eps_lim_proof}. 

\begin{figure}[h!]
\centering
\subfloat[$V=B$]{
\includegraphics[trim = 0 0 1.5mm 2.2mm, width=0.5\linewidth, clip]{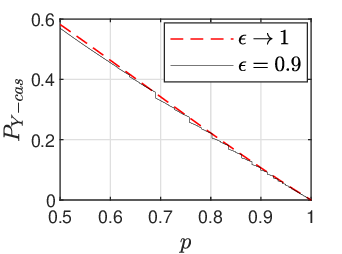}
\label{eps_lim_VB}}
\subfloat[$V=G$]{
\includegraphics[trim = 0 0 1.5mm 2.2mm, width=0.5\linewidth, clip]{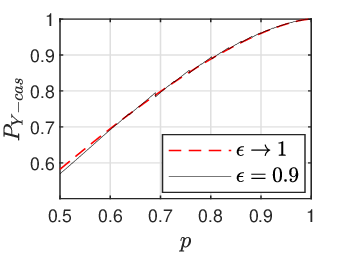}
\label{eps_lim_VG}}
\caption{\small Probability of $Y$ cascade versus private signal quality for the indicated values of $\epsilon$ under (a) $V=B$ and (b) $V=G$.}
\label{eps_lim_fig}
\end{figure}

Figure \ref{eps_lim_fig} illustrates \eqref{eps_lim_expression} and the corresponding probability when $\epsilon = 0.9$ as a function of the signal quality. For both $V=B$ and $V=G$, a better signal quality leads to improved learning even when fake agents have overwhelmed the ordinary agents. Also note that for $V=B$, for a weak signal quality, the incorrect cascade is more likely than the correct one (while for $V=G$, this is never true). 

\vspace{0.1mm}
The asymptotic welfare in this limiting scenario, $ \lim_{\epsilon \rightarrow 1 } \Pi(\epsilon)$ can be obtained by taking the limit $\epsilon \rightarrow 1$ in \eqref{Pi_expression} and then substituting \eqref{eps_lim_expression}. In Figure \ref{Exp-pay-off_fig}, observe that the sequences \scalebox{0.9}{$\{ \Pi (\epsilon_r^{-}) \}$} and \scalebox{0.9}{$\{ \Pi (\epsilon_r^{+}) \}$}, marked by $\times$ and $\circ$ respectively, converge to this limiting value as $r \rightarrow \infty$. Further, it can be proved that for any $p$, $ \lim_{\epsilon \rightarrow 1 } \Pi(\epsilon) > F$, where we recall that $F$ (defined in \eqref{F}) is the welfare when an agent acts based on its private signal alone. Therefore, even when fake agents have corrupted almost all the actions, it is still better for any ordinary agent to observe its past in addition to its private signal. Figure \ref{fig:welfare_increase_limeps1} plots the fractional improvement in asymptotic welfare relative to $F$ against varying values of $p$ and shows that it decays to $0$ with increasing $\mspace{1.5mu}p$.

\vspace{-3mm}
\section{Effects of a Platform Co-ordinator}
\label{sec:platform_coord}
In this section, we introduce an additional entity called the \emph{Platform Co-ordinator}, which at each time $i$, randomly modifies (manipulates or filters) the observation $O_i$ before presenting it to future agents. We assume that the Co-ordinator has the same information about private signal quality $p$ and fraction of fake agents $\epsilon$ as the rational buyers, while the item's underlying true value $(V)$ is unknown to both and assumed to be equiprobable. Moreover, the type of modification and the corresponding parameter(s) used by the Co-ordinator are common knowledge. By modifying the observations, the Co-ordinator who acts as a Bayesian \emph{persuader} aims to improve the asymptotic welfare. A related work is \cite{itai}, where a similar entity instead designs the information structure for the agents' private signals so as to maximize its utility which depends on the eventual learning outcome. We investigate the effects on agents' asymptotic welfare under three scenarios, namely wherein the Co-ordinator $(a)$ modifies a $N$ to a $Y$ w.p. $\kappa$ or $(b)$ discards a $Y$ w.p. $e$ or $(c)$ modifies a $Y$ to a $N$ w.p. $\beta$. Otherwise, in all scenarios, observations that are neither discarded nor modified are retained. Let ${}^{(i)}\Pi $ refer to the asymptotic welfare, where $i \in \{a,b,c\}$ indicates the particular scenario being considered. The superscript $(i)$ is dropped when referring to the asymptotic welfare without the Co-ordinator. 

\vspace{-3mm}
\subsection*{Scenario $\bm{(a)}$: Manipulating the \bm{$N$}'s}
 
Here, the Co-ordinator modifies only the $N$'s in the observations, i.e., at any time $i$, if $O_i = N$ then the modified observation $O_i^{'} = Y$ w.p. $\kappa \in [0,1)$ and $O_i^{'} = N$ w.p. $1 - \kappa$. Whereas, if $O_i = Y$ then $O_i^{'} = Y$ w.p. $1$. Here, $\mathcal{H}_{n-1}^{'} := \{O_1^{'},\ldots, O_{n-1}^{'} \}$ denotes the modified history observed by agent $n$. This has the effect that agents now perceive their history as having an effectively increased probability, $\epsilon_{\text{eff}} := \epsilon + (1-\epsilon)\kappa$ of past agents being fake. As a result, the asymptotic welfare in this scenario denoted by ${}^{(a)}\Pi(\epsilon, \kappa)$ is given by ${}^{(a)}\Pi(\epsilon, \kappa) = \Pi(\epsilon_{\text{eff}})$, where $\Pi(\cdot)$ is defined in \eqref{Pi_inf}. Note that by varying $\kappa$, $\epsilon_{\text{eff}}$ can attain any desired value in $[\epsilon,1)$. Specifically, to get the best improvement in asymptotic welfare in this scenario, the Co-ordinator can set $\kappa$ to the optimal value:
\begin{align}
\kappa^* = \underset{\kappa \in [0,1)}{\arg \max} \; \Pi \big( \epsilon + (1-\epsilon)\kappa \big),
\end{align} 
which yields the optimal asymptotic welfare,
\begin{align}
{}^{(a)}\Pi(\epsilon, \kappa^*) = \Pi\big( \epsilon + (1-\epsilon)\kappa^* \big).
\label{scene_a}
\end{align}

\vspace{-3mm}
\subsection*{Scenario \bm{$(b)$}: Filtering out the \bm{$Y$}'s }

\begin{figure}
\centering
\subfloat[]{
\begin{tikzpicture}[scale=1.27] 
\draw [decoration={markings,mark=at position 1 with {\arrow[scale=2,>=stealth]{>}}},postaction={decorate}] (0,0) -- (1.5,0.4); 
\node at (1.9,0.4) {\;\, \scalebox{0.95}{Discard}};
\draw [decoration={markings,mark=at position 1 with {\arrow[scale=2,>=stealth]{>}}},postaction={decorate}] (0,0) -- (1.5,-0.4);
\node at (1.63,-0.4) {$Y$};
\draw [decoration={markings,mark=at position 1 with {\arrow[scale=2,>=stealth]{>}}},postaction={decorate}] (0,-1) -- (1.5,-1);
\node at (1.63,-1) {$N$};
\node at (-0.1,0) {$Y$};
\node at (-0.1,-1) {$N$};
\node at (-0.3,-0.5) {$O_i$};
\node at (2.1,-0.5) {$O_i^{'}$};
\node at (0.75,0.35) [rotate= 15] {\scalebox{0.9}{$e$}};
\node at (0.75,-1.15) {\scalebox{0.9}{$1$}};
\node at (0.8,-0.4) [rotate= -15] {\scalebox{0.9}{$1 - e$}};
\end{tikzpicture}
\label{Filter_channel}}
\subfloat[]{
\includegraphics[trim = 0 0 0 2.5mm, width=0.57\linewidth, clip]{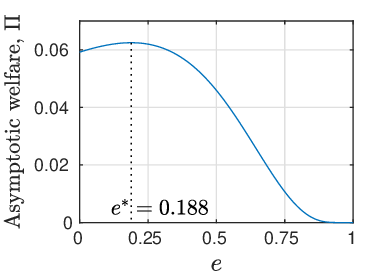}  
\label{jump_size_figure}}
\setlength{\abovecaptionskip}{7pt}
\setlength{\belowcaptionskip}{-4pt}
\caption{\small$(a)$ The channel through which observations are filtered by the Co-ordinator at each time $i$. $(b)$ Asymptotic welfare $^{(b)}\Pi(\epsilon,e)$ as a function of the filtration parameter $e$, for $p=0.6$ and $\epsilon = 0.9$. Maximal value obtained at $e = e^*$.}
\vspace{-3mm}
\end{figure}

In this scenario, the Co-ordinator randomly filters out the $Y$'s from the observations instead of manipulating the observations as was done in Scenario $(a)$. Specifically, at every time $n$, the Co-ordinator discards a $Y$ observation w.p. $e \in [0,1)$ and retains it otherwise. Whereas, a $N$ observation is always retained. This bias is due to the fact that, while a $N$ observation always represents the action of a rational agent, a $Y$ observation could also represent the action of a fake agent. So, filtering out some of the $Y$ observations while retaining all the $N$ observations seems reasonable. The channel in Figure \ref{Filter_channel} depicts the filtering of observation $O_i$ at each time $i$, where $O_i^{'} \in \{Y,N, \text{Discard}\} $ denotes the channel output. Here, $O_i^{'} = \text{Discard}$ implies that a $Y$ at time $i$ has been discarded by the Co-ordinator.  

\vspace{0.1mm}
Note that the arrival of an agent into the platform is recorded for (and visible to) subsequent agents only if its corresponding observation is retained by the Co-ordinator. If an agent's observation is discarded, future agents are unaware of its arrival into the buying platform. Therefore, we define a new index set $\{i_1, i_2, \ldots\} \subseteq \{1,2,\ldots\}$ which indexes the arrivals of only those agents whose observations are retained. Here, $i_n$ is the index in the unfiltered arrival sequence of the agent whose observation is the $n^{\text{th}}$ undiscarded observation.

\vspace{0.2mm}
Now, for the $i_n^{\text{th}}$ agent which observes a history $\mathcal{H}_{i_{n-1}} := \{O_{i_1}, \ldots, O_{i_{n-1}}\}$, the Bayes' optimal decision rule is still given by \eqref{bayes_decison}, except that the posterior probability for the item being good, $\gamma_{i_n} := \mathbb{P} (G \vert S_{i_n}, \mathcal{H}_{i_{n-1}} )$. Further, with the likelihood ratio for its history $\mathcal{H}_{i_{n-1}}$ being redefined as $l_{i_{n-1}} \! \triangleq \! \mathbb{P}(\mathcal{H}_{i_{n-1}} \vert B) / \mathbb{P}(\mathcal{H}_{i_{n-1}} \vert G) $, it can be shown that both Lemma \ref{lemma1} and Property \ref{prop1} remain true. Now, assuming the $i_n^{\text{th}}$ agent does not cascade, Property \ref{prop1} and Lemma \ref{lemma1} imply that for every observation $O_{i_k}, k \leq n$, $A_{i_k}$ follows $S_{i_k}$. Then, for every  such observation $O_{i_k}$, the probability that it follows the true value $V$ is implicitly conditioned on the fact that it has not been discarded, and hence is given by 

\vspace{-0.5mm}
{ \small
\setlength{\abovedisplayskip}{-1pt}
\begin{align}
\hspace{-2mm} a^{'} &= \mathbb{P}(O_{i_k} = Y \vert \, G, O_{i_k} \neq \text{Discard}) = \frac{a(1-e)}{a(1-e) + (1-a)}, \label{a_prime} \\
\hspace{-2mm} b^{'} &= \mathbb{P}(O_{i_k} = N \vert \, B, O_{i_k} \neq \text{Discard}) = \frac{b}{b + (1-b)(1-e)}.  \label{b_prime}
\end{align}}%
Further, for every such $O_{i_k}$, as $S_{i_k}$ is conditionally independent of the history $\mathcal{H}_{i_{k-1}}$ given $V,$ the likelihood ratio updates are $l_{i_k} =$\scalebox{0.92}{$\Big( \frac{1-b}{a} \Big)$}$l_{i_{k-1}}$ if $O_{i_k} = Y$, and $l_{i_k} =$\scalebox{0.92}{$\Big( \frac{b}{1-a} \Big)$}$l_{i_{k-1}}$ if $O_{i_k} = N$. These updates are identical to the likelihood updates in \eqref{likelihood_update}, which are for the original model without the Co-ordinator. This is because, such an observation $O_{i_k}$, which could be an undiscarded $Y$ or an $N$, still possesses the same information about $V$ as in the original model. Now, as a result of the updates, $l_{i_n}$ can be expressed as $l_{i_n} = \big( {\small \text{$\frac{1-p}{p}$}} \big)^{ \scriptsize \text{$h_{i_n}$}}$ where $h_{i_n}$ is a sufficient statistic of the information contained in history $\mathcal{H}_{i_n}$, given that agent $i_n$ is not in a cascade. Similar to eq. \eqref{h_n} with $\eta$ as per \eqref{eta}, 
{\setlength{\abovedisplayskip}{4.2pt}
\setlength{\belowdisplayskip}{4.2pt}
\begin{align}
h_{i_n} := \eta n_Y - n_N,   \label{h_n_prime}
\end{align}}%
which is the difference between the number of $Y$'s $(n_Y)$ weighted by $\eta$ and the number of $N$'s $(n_N)$ present in $\mathcal{H}_{i_n}$. 

\vspace{0.1mm}
Now, Lemma \ref{lemma1} implies that until a cascade occurs, $-1 \leq h_{i_n} \leq 1$ for all such times $n$, and \eqref{h_n_prime} implies that the update rule for $h_{i_n}$ is given by
\begin{align}
h_{i_n} = \begin{cases} h_{i_{n-1}} + \eta \quad & \text{if} \;\; O_{i_n} = Y,\\
h_{i_{n-1}} - 1  \quad & \text{if} \;\; O_{i_n} = N.
\end{cases}
\label{mc_update_two_fake}
\end{align}
Once $h_{i_n} >1 $ ($ < -1 $), a $Y \,(N)$ cascade begins and $h_{i_n}$ stops updating (Property \ref{prop1}). Observe that the updates of $ \{ h_{i_n} \}$ stated in \eqref{mc_update_two_fake} are identical to the updates of $\{h_n\}$ in \eqref{h_n}. However, $ \{ h_{i_n} \}$ has two notable differences. Firstly, $ \{ h_{i_n} \}$ is indexed by the new set $\{i_1, i_2, \ldots\}$ and secondly, conditioned on $V$, $ \{ h_{i_n} \}$  moves to the right by $\eta$ w.p. $\mathbb{P}(O_{i_n} = Y \vert V)$ or to the left by $1$ w.p. $\mathbb{P}(O_{i_n} = N \vert V) $ until a cascade occurs, where these probabilities are defined in terms of $a^{'}$ and $b^{'}$ in \eqref{a_prime} and \eqref{b_prime}. This is unlike in the case of $\{h_n\}$ where $a$ and $b$ in \eqref{a_b_expressions} define the transition probabilities.

\vspace{0.2mm}
Next, let the $Y$ cascade probability for the process  $\{h_{i_n}\}$ be denoted by ${}^{(b)}\mathbb{P}_{Y\text{-cas}}^V(\epsilon, e)$ to highlight that it is a function of the fraction of fake agents $\epsilon$ and the filtration parameter $e$ set by the Co-ordinator. Similarly, let ${}^{(b)}\Pi(\epsilon, e)$ denote the asymptotic welfare associated with this scenario, which relates to the cascade probablities as per \eqref{Pi_expression} as
\begin{align}
^{(b)}\Pi(\epsilon,e) &=  (1/4) \big[ \, ^{(b)}\mathbb{P}_{Y\text{-cas}}^G (\epsilon,e) - ^{(b)}\mathbb{P}_{Y\text{-cas}}^B (\epsilon,e) \big].   \label{Pi_scene_b}
\end{align}
Now, as both processes: $\!\! \{h_{i_n}\}$ and $\{h_n\}$ share the same update rule, \scalebox{0.95}{${}^{(b)}\mathbb{P}_{Y\text{-cas}}^V(\epsilon, e)$} can be computed using the recursive method described in Section \ref{sec:markov} by equations \eqref{reccur} and \eqref{PYcas}, except that $p_f = a^{'}$ if $V=G$, else $p_f = 1-b^{'}$ if $V=B$. Then, applying these probabilities to \eqref{Pi_scene_b} yields $^{(b)}\Pi(\epsilon,e)$. For the best improvement in asymptotic welfare, the Co-ordinator needs to set $e$ to the optimal value:
\begin{align}
e^* = \underset{e \in [0,1)}{\arg \max} \; {}^{(b)}\Pi(\epsilon, e),
\end{align} 
which yields the best welfare, ${}^{(b)}\Pi(\epsilon, e^*)$. As an example, Figure \ref{jump_size_figure} plots this welfare against $e$ for $p=0.6$ and $\epsilon = 0.9$, where the best welfare is obtained at $e^* = 0.188$. 


\subsection*{Scenario $\bm{(c)}$: Manipulating the \bm{$Y$}'s}
Here, as opposed to Scenario $(a)$, the Co-ordinator modifies only the $Y$'s in the observations, i.e., at any time $i$, if $O_i = Y$ then the modified observation $O_i^{'} = N$ w.p. $\beta \in (0,1)$ and $O_i^{'} =Y$ w.p. $1 - \beta$. Whereas, if $O_i = N$ then $O_i^{'} = N$ w.p. $1$. The net effect of these random modifications is that, at any time $i$, the channel between action $A_i$ and the modified observation $O_i^{'}$ becomes a binary asymmetric channel with cross-over probabilites: 
\begin{align}
\epsilon_Y &:= \mathbb{P} (O_i^{'} = Y \vert A_i =N ) = \epsilon (1-\beta) \nonumber \\
\text{and} \;\; \epsilon_N &:= \mathbb{P} (O_i^{'} = N \vert A_i =Y ) = \beta.
\label{epsY_epsN_(c)}
\end{align}
For this scenario, let the $Y$-cascade probability be denoted by ${}^{(c)}\mathbb{P}_{Y\text{-cas}}^V(\epsilon, \beta)$ to highlight that it is a function of the fraction of fake agents $\epsilon$ and the modification parameter $\beta$. Similarly, let ${}^{(c)}\Pi \left( \epsilon, \beta \right) $ denote the asymptotic welfare associated with this scenario. It is difficult to compute the $Y$ cascade probabilites for general values of the channel parameters: $\epsilon_Y \in (0,\epsilon)$ and $\epsilon_N = (0,1)$, except when $\epsilon_Y = \epsilon_N$. This occurs if the Co-ordinator sets $\beta = \beta_{\text{sym}} := \frac{\epsilon}{1+\epsilon}$ at which the channel between $A_i$ and $O_i^{'}$ effectively becomes a BSC with cross-over probability $\beta_{\text{sym}}$. In Appendix \ref{app:scene_c}, we show that for this special case, a sufficient statistic of history $\mathcal{H}_n^{'}$, at each time $n$ is a random walk $\{s_n\}$ that occupies a finite state-space (unlike $\{h_n\}$ in \eqref{h_n} that typically has a countably infinite state-space). Thus, closed-form expressions for cascade probabilities exist and are derived in Appendix \ref{app:scene_c}. The resultant asymptotic welfare is given by
\begin{align}
^{(c)} \Pi (\epsilon, \beta_{\text{sym}}) =  (1/4) \,  \frac{a_{\text{sym}}^{k} - (1-a_{\text{sym}})^{k}}{a_{\text{sym}}^{k} + (1-a_{\text{sym}})^{k}}.
\label{Pi_scene_c}
\end{align}
where $ a_{\text{sym}} := p(1-\beta_{\text{sym}}) +(1-p)\beta_{\text{sym}}$ and $k := \left \lfloor  \log_{ \text{\scalebox{0.92}{$ (1-a_{\text{sym}}) / a_{\text{sym}} $}} } \big( \text{\scalebox{0.87}{$\frac{1-p}{p}$}} \big) \right \rfloor +1.$

\vspace{-2mm}
\subsection*{Asymptotic welfare comparisons}
\begin{figure}[h!]
\centering
\includegraphics[trim = 0cm 1mm 0cm 4.5mm, width=\linewidth, clip]{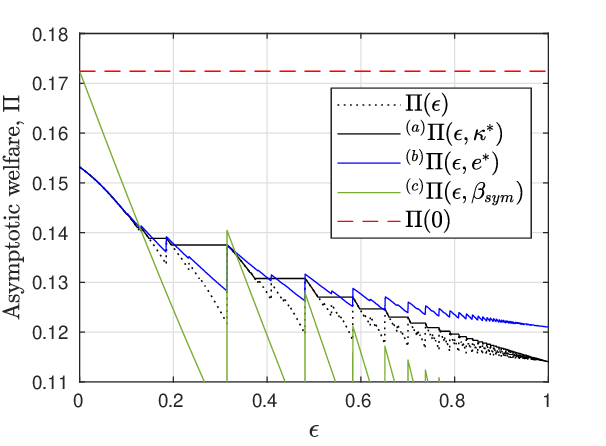}
\setlength{\abovecaptionskip}{-5pt}
\setlength{\belowcaptionskip}{-6pt}
\caption{\small Asymptotic welfare for the indicated scenarios versus the fraction of fake agents for $p=0.7$.}
\label{fig:welfare_compare}
\end{figure}

We now compare the three asymptotic welfares, namely $^{(a)}\Pi(\epsilon, \kappa^*)$ , $^{(b)}\Pi(\epsilon, e^*)$ and $^{(c)} \Pi (\epsilon, \beta_{\text{sym}})$ for fixed values of the private signal quality $p$. We also contrast these welfares with the default welfare in the absence of the Co-ordinator, $\Pi(\epsilon)$ and the baseline welfare $\Pi(0)$ (given by \eqref{Pi_inf} and \eqref{Pi_zero} resp.). Figure \ref{fig:welfare_compare} plots the different welfares against $\epsilon$ for $p=0.7$. It can be seen that for low values of $\epsilon$, Scenario $(c)$ provides the best improvement in welfare. We also observe that as $\epsilon \rightarrow 0$, Scenario $(c)$ entirely mitigates the reduction in welfare caused by the presence of fake agents. For high values of $\epsilon$, Scenario $(b)$ provides the best improvement in welfare. Here, Scenario $(c)$ in fact significantly worsens the welfare as compared to $\Pi(\epsilon)$. There also exist several intervals in $[0,1)$ with moderate values of $\epsilon$ where Scenario $(a)$ performs the best. We find the above characteristics of the three scenarios to be consistent for all values of $p$. We demonstrate this in Appendix \ref{app:scenario_plots} with plots for two more values, $p=0.55$ and $p=0.9$.

\vspace{-3mm}
\section{General case of zero ex ante pay-off}  \label{sec:general_prior}
In this section, we extend our model to consider a general (possibly non-uniform) prior for the true value $V$ of the item and a general pay-off structure for the agents, while still retaining the condition of zero \emph{ex-ante} pay-off. Recall that this condition implies agents' indifference to the two actions a priori. We begin by assuming a non-revealing general prior for the true value, $\mathbb{P}(V=G) = q \in (0,1)$. Recall form Section \ref{sec:model} that in general, the amount a buyer gains (loses) is $x$ $(y)$ if $V = G$ $(V = B)$, where $x,y \geq 0$ and $C$ denotes the cost of buying the item. Restricting the \emph{ex-ante} pay-off to $0$ implies that $C = qx - (1-q)y$. Then, the general pay-off for any agent $i$ is given by

{ \setlength{\abovedisplayskip}{-5pt}
\setlength{\belowdisplayskip}{7pt}
\begin{align}
\mspace{-10mu}  \pi_i = \begin{cases}   \text{\scalebox{0.93}{$x-C =   (1-q)(x+y),$}} \!  &\text{if} \; \text{\scalebox{0.95}{$A_i=Y$}} \; \text{and} \; \text{\scalebox{0.95}{$V=G,$}} \\
\text{\scalebox{0.93}{$-y-C = -q(x+y),$}} \! &\text{if} \; \text{\scalebox{0.95}{$A_i=Y$}} \; \text{and} \; \text{\scalebox{0.95}{$V=B,$}} \\
0,  &\text{if} \; \text{\scalebox{0.95}{$A_i=N.$}} \\
\end{cases}
\end{align}}%
In all previous sections, we have considered agents' pay-offs under the specific case of $x = 1$, $y=0$ and $C = 1/2$. 

\vspace{0.06mm}
In this more general scenario, the decision rule is similar to \eqref{bayes_decison} except $\gamma_n = q$ is the new threshold at which agent $n$ is indifferent to the two actions (and thus follows $S_n$). Next, using Bayes' rule, we express $\gamma_n$ in terms of $l_{n-1}$ and $\beta_n$ as $\gamma_n = 1/(1+\beta_n l_{n-1} \frac{1-q}{q})$. Then, the condition on $\gamma_n$ for a $Y \, (N)$ cascade is $\gamma_n >q $ $(<q)$ for all $S_n$, which translates to $\beta_n l_{n-1} <1$ $(>1)$ for all $S_n$. This cascade condition is the same as before (see proof of Lemma \ref{lemma1}) and hence is still defined in terms of $l_{n-1}$ as per Lemma \ref{lemma1}. As a result, the sequence of agents' actions once again satisfies Properties \ref{prop1} and \ref{prop2} where  it is governed by $\{h_n\}$ which starts in the state $h_0=0$,\footnote{If the ex-ante pay-off is not restricted to $0$, then for an arbitrary cost $C \in [-y,x]$, the resulting r.w. $\{h_n\}$ still updates as per \eqref{mc_update}, except that the starting state $h_0 = \log \left( \frac{q(x-C)}{(1-q)(y+C)} \right) / \log \left( \frac{p}{1-p} \right)$.} evolves as per the update rule in \eqref{mc_update}, and is depicted in Figure \ref{random_walk}. Now, conditioned on the true value $V,$ $\{h_n\}$ does not change with the prior $q$. Thus, all  conditional probabilities derived in this paper remain unaltered in the general scenario. On the contrary, agents' welfares being their pay-offs averaged over $V$ with a general prior $q$ will now change as follows. Evaluating \eqref{Exp_pi_n} given that agent follows its private signal yields a new  $F := q(1-q)(x+y)(2p-1)$. Next, the constant term $1/4$ is replaced by the term $q(1-q)(x+y)$ in all equations pertaining to agents' welfares, namely, equations \eqref{E_pi_n}, \eqref{Pi_inf}, \eqref{Pi_zero}, \eqref{Delta_r}, \eqref{Pi_eps_lim_zero}, \eqref{Pi_scene_b}, \eqref{Pi_scene_c}. However, in Eq. \eqref{Pi_inf}, Step $(a)$ is no longer true and $\mathbb{P}_{\text{wrong-cas}} := q \mathbb{P}_{N\text{-cas}}^G + (1-q) \mathbb{P}_{Y\text{-cas}}^B.$  With the above changes in place, all results and discussions presented in this paper extend to any case with zero ex ante pay-off. Our analytic techniques can also be readily modified for non-zero ex-ante pay-off, as done in \cite{pawanWiopt2023}.  

\vspace{-3mm}
\section{Conclusions and future work} \label{concl}
We studied the effect of randomly arriving fake agents, who by taking a fixed action seek to influence the outcome of an information cascade. We focussed on the impact of varying the fraction of fake agents on the probability of their preferred cascade. To study this impact, we developed a Markov chain model which typically has a countably infinite state-space and does not readily allow for a closed-from solution to the cascade probabilities. Instead, we presented an iterative method that can compute the cascade probabilities with arbitrary precision. This process also yields exact values for any given agent’s chances of herding and its welfare, which is the expected pay-off it receives.

Our main result identified scenarios where surprisingly, an increase in the fraction of fake agents not only reduces the chances of their preferred cascade but also effects a significant improvement in the welfare of every rational agent. Further, we analysed three approaches to modify the observation database such that learning (asymptotic welfare) can be improved, namely, $(1)$ increasing, $(2)$ filtering out and $(3)$ modifying the possibly fake actions. Interestingly, we observed that the third approach provides the best improvement in learning when the original fraction of fake agents is low, with a complete mitigation of their welfare-reducing effects as their fraction tends to zero. Whereas, the second approach performs the best for high values of this fraction. Lastly, we showed that our analysis, results and discussions readily extend to general priors and agent pay-offs, as long as the ex-ante pay-off is zero. For non-zero ex-ante pay-off, our analytic techniques can also be easily modified to study the platform's behaviour. 


Potential future directions for this work include studying the effects of time-varying fractions and/or multiple types of fake agents, non-Bayesian rationality, random and asymmetric tie-breaking rules, and imperfect observations such as allowing each agent to only observe the total number for each action-type in its history, instead of perfectly observing the sequence of past actions. Moreover, as our results rely on the probability of fake agents being common knowledge, a natural extension would be to relax this assumption.

\vspace{-0.3mm}

\bibliographystyle{IEEEtran}


\clearpage

\appendices

\section{Proof of Theorem \ref{Thm:1}} \label{app:Thm1_proof}

\begin{IEEEproof}
Recall that \scalebox{0.95}{$P_{M+1}=1$} while computing \scalebox{0.95}{$\mathbb{P}_{Y\text{-cas}}^M$}, which implies that \scalebox{0.95}{$\mathbb{P}_{Y\text{-cas}}^M$} is the net probability of $(a)$ sequences that terminate in a $Y$ cascade by the $M^{\text{th}}$ stage and $(b)$ sequences that do not terminate by the $M^{\text{th}}$ stage. Thus, the difference: \scalebox{0.9}{$\mathbb{P}_{Y\text{-cas}}^M - \mathbb{P}_{Y\text{-cas}}$} is upper-bounded by the probability of $(b)$. Accounting for all possible sequence combinations that persist through stages \scalebox{0.95}{$1,2,\ldots,M$} yields the probability of $(b)$:
\begin{align*}
\prod_{i=1}^{M} \big[ r_i (1-p_f) p_f^{r_i} \big] \leq \big[ (r+1)(1-p_f) p_f^r \big]^M = k^M
\end{align*}
where $k$ is as defined in Theorem \ref{Thm:1}. Next, for a fixed $p$ and for $\epsilon \in \mathcal{I}_r$, $k$ is maximized only if $\epsilon$ is such that $p_f = \frac{r}{1+r}$, which may not be satisfied for any $\epsilon$ in $\mathcal{I}_r$. Nevertheless, assuming $k$ is maximized, the maximal value of $k$ would be $ \big( \frac{r}{1+r} \big)^r $. As this maximal value decreases in $r$, evaluating it at $ r=1$ yields \scalebox{0.95}{$k \leq 1/2$} for any $r$.
\end{IEEEproof}

\vspace{-2mm}
\section{} \label{app:PYcas_VGood}

\section{$\bm{N}$ cascade probability, $\bm{\mathbb{P}_{N\text{-cas}}}$}
\label{app:PNcas}

In the iterative method of Figure \ref{iter_struct_Ncas}, with $ \epsilon \in \mathcal{I}_r \smallsetminus \{\epsilon : 1/\eta \in \mathbb{Q}\}$, it can be shown that the successive values of $r_i$ for $i=1,2,3,\ldots$ are exactly same as those defined in \eqref{r_updates} for the $Y$ cascade scenario, with $r_1 = r+1$.


\begin{figure}[h]  
\centering
\begin{tikzpicture}[scale=2]
\begin{scope}[shift={(-3,0)}]
\draw [dashed, decoration={markings,mark=at position 1 with {\arrow[scale=1,>=stealth]{>}}},postaction={decorate}] (0.2,0) -- (0.85,0) -- (0.85,0.2); 
\node at (0.7,0.3) {\scalebox{0.8}{$N$ cascade}};
\node at (-0.04,0) {$N \; N$}; 
\node at (0,-0.2) {$2 \, N \; Y \; N$}; 
\draw [dashed] (0.4,-0.2) -- (0.85,-0.2) -- (0.85,0);
\draw [dashed] (0.4,-0.4) -- (0.85,-0.4) -- (0.85,-0.2);
\node at (0.02,-0.4) {$3 \, N \; Y^2  N$};  
\draw [dotted] (0,-0.5) -- (0,-0.9);
\node at (0.2,-1.0) {$r_1 \, N \; Y^{r_1 -1} N$};
\draw [dashed , rounded corners] (0.8,-1.0) -- (0.85,-1.0)--(0.85,-0.4); 
\node at (0.02,-1.2) {$r_1 \, N \; Y^{r_1}$}; 
\draw [rounded corners, decoration={markings,mark=at position 1 with {\arrow[scale=2,>=stealth]{>}}},postaction={decorate}] (0.4,-1.2) -- (1.05,-1.2); 
\draw [dashed] (-0.45,0.1) -- (-0.45,-1.3);
\draw [dashed] (1.05,0.1) -- (1.05,-1.3);
\node at (0.23,-1.7) {Stage $(1)$};
\end{scope}
\begin{scope}[shift={(-1.5,0)}]
\draw [dashed, decoration={markings,mark=at position 1 with {\arrow[scale=1,>=stealth]{>}}},postaction={decorate}] (0.2,0) -- (0.85,0) -- (0.85,0.2); 
\node at (0.7,0.3) {\scalebox{0.8}{$N$ cascade}};
\node at (-0.04,0) {$N \; N$}; 
\node at (0,-0.2) {$2 \, N \; Y \; N$}; 
\draw [dashed] (0.4,-0.2) -- (0.85,-0.2) -- (0.85,0);
\draw [dashed] (0.4,-0.4) -- (0.85,-0.4) -- (0.85,-0.2);
\node at (0.02,-0.4) {$3 \, N \; Y^2  N$};  
\draw [dotted] (0,-0.5) -- (0,-0.9);
\node at (0.2,-1.0) {$r_2 \, N \; Y^{r_2 -1} N$};
\draw [dashed , rounded corners] (0.8,-1.0) -- (0.85,-1.0)--(0.85,-0.4); 
\node at (0.02,-1.2) {$r_2 \, N \; Y^{r_2}$}; 
\draw [rounded corners, decoration={markings,mark=at position 1 with {\arrow[scale=2,>=stealth]{>}}},postaction={decorate}] (0.4,-1.2) -- (1.05,-1.2); 
\draw [dashed] (-0.45,0.1) -- (-0.45,-1.3);
\draw [dashed] (1.05,0.1) -- (1.05,-1.3);
\node at (0.23,-1.7) {Stage $(2)$};
\node at (1.3,-0.6) {$\ldots$};
\end{scope}
\end{tikzpicture}
\caption{\small An enumeration of all possible sequences that would lead to a $N$ cascade. The term $Y^{t}$ represents $t$ consecutive $Y$'s and the term  $(t) \, NY^{t-1}N$ represents the $t$ allowable permutations of the sub-sequence $NY^{t-1}N$.}  
\label{iter_struct_Ncas}
\end{figure} 

\noindent Further, let $T_n$ denote the probability of getting absorbed into a $N$ cascade given that the sequence has not terminated before the $n^{\text{th}}$ stage. Then, the following recursion holds:
{\small
\begin{align}
T_n = (1-p_f)^2 \sum_{t=1}^{r_n} t p_f^{t-1} + r_n(1-p_f) p_f^{r_n} T_{n+1}, \;\; \text{for} \; n=1,2,\ldots 
\label{reccurNcas}
\end{align}}%
and the probability of a $N$ cascade, denoted by $\mathbb{P}_{N\text{-cas}}$ is
\begin{align}
\mathbb{P}_{N\text{-cas}} (\epsilon) = T_1, \quad \text{for} \;\; \epsilon \in \mathcal{I}_r \smallsetminus \{\epsilon : 1/\eta \in \mathbb{Q}\}.
\label{PNcas}
\end{align}
Equations \eqref{reccurNcas}, \eqref{PNcas} and \eqref{r_updates} outline the iterative method to compute $\mathbb{P}_{N\text{-cas}}$, where in practice we again truncate the process to a finite number of iterations $M$ by  assuming $T_{M+1}=1$. The resulting value: $\mathbb{P}_{N\text{-cas}}^M$ can be shown to be a tight upper bound to $\mathbb{P}_{N\text{-cas}}$ as $M \rightarrow \infty$ by using techniques very similar to the proof of Theorem \ref{Thm:1}.

\section{Proof of Lemma \ref{lemma_for_ri}}  \label{app:lemma_ri_proof}

\begin{IEEEproof}
We consider two cases: $\epsilon \rightarrow \epsilon_r^+$ and $\epsilon \rightarrow \epsilon_r^-$. The proof outline for the first case is as follows: assume $r_j = r$ for all $ 2 \leq j \leq i-1$ and note that $r_1=r+1$ as $\epsilon_r^+ \in \mathcal{I}_r$. Then it follows from \eqref{r_updates} that $r_i = r $ only if $\eta > \big( r+ \frac{1}{i} \big)^{-1}$. Now as $\epsilon \rightarrow \epsilon_r^+$, $\eta \rightarrow 1/r$ and hence this condition is satisfied. Using this argument inductively shows that $r_i =r$ for all $ i \geq 2$. The second case is proved similarly.
\end{IEEEproof}

\section{Proof of Theorem \ref{thm:welfare_agent_n_@eps_r}}   \label{app:welfare_@eps_r}
Given true value $V \in \{G,B\}$, for any agent index $i \in \mathbb{N}$ and at any $r^{\text{th}}$ $\epsilon$-threshold $\epsilon_r$, $r \geq 2$, we define:
\begin{align}
\Delta v_i^V := v_i^V (\epsilon_r^{+}) - & v_i^V (\epsilon_r^{-}), \;\;\;  \Delta u_i^V := u_i^V (\epsilon_r^{+}) - u_i^V (\epsilon_r^{-}),  \nonumber \\
d_i &= (1-p) \Delta v_i^G - p \Delta v_i^B  \;\; \text{and} \label{d_i} \\
f_i &= (1-p) \Delta u_i^B - p \Delta u_i^G.  \label{f_i}
\end{align}
Then, by applying the expression for $E[\pi_n]$ given in \eqref{E_pi_n}, we have for any agent $n $:
\begin{align}
\mathbb{E} [ \pi_n(\epsilon_r^+)] - \mathbb{E} [ \pi_n(\epsilon_r^-)] =  \sum_{i=1}^n ( d_i + f_i ).
\label{d_i_plus_f_i_master}
\end{align}
To prove Theorem \ref{thm:welfare_agent_n_@eps_r} is to show that the above quantity is non-negative for all $n$. To that end, we first present Lemmas \ref{lemma4} and \ref{lemma5} which provide expressions for the sequences $\{d_i\} $ and $\{f_i\}$.


As a preface to the lemmas, note that the quantities $v_i^V$ and $u_i^V$ defined in \eqref{v_n} and \eqref{u_n} are expressed with respect to the sequence $\{r_i\}$ defined in \eqref{r_updates}. As per Lemma \ref{lemma_for_ri}, $r_i = r$ for every $i \geq 2$ for both $\epsilon = \epsilon_r^+$ and $\epsilon_r^{-}$. However, the value of $r_1$ differs and it equals $r+1$ for $\epsilon_r^+$ and $r$ for $\epsilon_r^-$. To avoid confusion, we set the value of $r_1$ to $r+1$, i.e., as per $\epsilon_r^+$ and make due adjustments when evaluating quantities at $\epsilon_r^-$. To express $v_i^V$ and $u_i^V$, we also require the special agent index $l_j$ for $j \in \mathbb{N}$ that is associated with each Stage $(j)$ of the enumeration processes of both $Y$ and $N$ cascades described in Figures \ref{iter_struct} and \ref{iter_struct_Ncas}. We restate it here for convenience.
\begin{align}
l_j \triangleq  r_j +\sum_{i=1}^{j-1} (r_i+1), \;\; \text{for}  \;\; j = 1,2,\ldots
\label{l_j}
\end{align}
Lastly, we will be using the below identities at many instances in this proof to simplify expressions.
\begin{align}
\frac{b}{1-a} &= \frac{p}{1-p} \quad \text{for any} \;\; \epsilon, \label{identity1}\\
\Big( \frac{a}{1-b} \Big)^r &= \frac{p}{1-p} \quad \text{at} \;\; \epsilon =  \epsilon_r, r \in \mathbb{N}. \label{identity2}
\end{align}
Here, \eqref{identity1} follows from definitions of $a$ and $b$ in \eqref{a_b_expressions} and \eqref{identity2} follows from fact that $\eta$ defined in \eqref{eta} equals $1/r$ when $\epsilon = \epsilon_r$, for any $r \in \mathbb{N}$.

\begin{lemma} For agent index $n$,
\begin{align}
d_n = \begin{cases}  \text{\scalebox{0.9}{$ p \left[ p_f^{r_j} \displaystyle \prod_{i=1}^{j-1} r_i (1-p_f) p_f^{r_i} \right]^B \! \! \! \!\! \left( \frac{a}{1-b} -1 \right) > 0,$}}  & \text{\scalebox{0.9}{if $n=l_j, j \in \mathbb{N},$}} \\
0, \quad &\text{o.w.} 
\end{cases} \label{eq0}
\end{align} \label{lemma4}
\end{lemma}

\begin{IEEEproof}
Observe the expression for $v_i$ in \eqref{v_n}. Note that $v_i^V (\epsilon_r^+)$ is non-zero only at $n = l_j$, whereas $v_i^V (\epsilon_r^-)$ is non-zero only at $n = l_j -1$. Therefore, $d_n$ defined in \eqref{d_i} is trivially zero at all indices except at $n=l_j$ and $l_j -1$. 

At $n = l_j -1$, $v_i^V (\epsilon_r^+) = 0$. Therefore, 
\begin{align}
d_n = p  v_i^B (\epsilon_r^{-}) \left[ 1 - \Big( \salphainv \Big) \frac{ v_i^G (\epsilon_r^{-})}{v_i^B (\epsilon_r^{-})} \right].  \label{eq1}
\end{align}
Now the ratio $v_i^G (\epsilon_r^{-}) / v_i^B (\epsilon_r^{-})$ can be simplified as follows:
\begin{align}
\frac{v_i^G (\epsilon_r^-)}{v_i^B (\epsilon_r^-)} &= \frac{ \left[ p_f^{r_j} \displaystyle \prod_{i=1}^{j-1} (r_i (1-p_f) p_f^{r_i}) \right]^G}{\left[ p_f^{r_j} \displaystyle \prod_{i=1}^{j-1} (r_i (1-p_f) p_f^{r_i}) \right]^B}  \;\; \text{with} \; r_1 = r \nonumber \\
 &= \left( \frac{1-a}{b} \right)^{j-1} \Big( \frac{a}{1-b} \Big)^{\displaystyle \sum_{i=1}^{j} r_i} \overset{(a)}{=} \frac{p}{1-p}. \label{eq2}
\end{align} 
Step $(a)$ follows by the use of identities \eqref{identity1} and \eqref{identity2}. Substituting the value obtained in \eqref{eq2} in \eqref{eq1} implies that $d_n = 0$ at $n = l_j -1$. Next, at $n = l_j$, $v_i^V (\epsilon_r^-) = 0$ which simplifies $d_n$ to
\begin{align}
d_n = p  v_i^B (\epsilon_r^{+}) \left[ \Big( \salphainv \Big) \frac{ v_i^G (\epsilon_r^{+})}{v_i^B (\epsilon_r^{+})} -1 \right],  \label{eq3}
\end{align}  
where the expression for $v_n^V (\epsilon_r^{+})$ at $n = l_j$ is given in \eqref{v_n}. Using this expression, the ratio $v_i^G (\epsilon_r^{+}) / v_i^B (\epsilon_r^{+})$ can be simplified as follows:
\begin{align}
\frac{v_i^G (\epsilon_r^{+})}{v_i^B (\epsilon_r^{+})} &=  \frac{ \left[ p_f^{r_j} \displaystyle \prod_{i=1}^{j-1} (r_i (1-p_f) p_f^{r_i}) \right]^G}{\left[ p_f^{r_j} \displaystyle \prod_{i=1}^{j-1} (r_i (1-p_f) p_f^{r_i}) \right]^B}  \;\; \text{with} \; r_1 = r+1 \nonumber \\
 &= \left( \frac{1-a}{b} \right)^{j-1} \Big( \frac{a}{1-b} \Big)^{\displaystyle \sum_{i=1}^{j} r_i} \overset{(b)}{=} \Big( \frac{p}{1-p} \Big) \Big( \frac{a}{1-b} \Big). \label{eq4}
\end{align}
Again, Step $(b)$ follows from identities \eqref{identity1}, \eqref{identity2}. Substituting the value of this ratio in \eqref{eq3} gives the required expression for $d_n$ at $n = l_j$, stated in Lemma \ref{lemma4}.
\end{IEEEproof}

\begin{lemma}
Given a fixed $j \in \mathbb{N}$ and $ t \in \{ 0,1,\ldots, r-1 \}$, for agent index $n = l_j + t +2$,
\begin{equation}
\begin{split}
f_n &= - \text{\scalebox{0.9}{$ \displaystyle  \left[ (1-p_f)^3 \big(r p_f^r (1-p_f) \big)^{j-1} p_f^{r+t} (r-t) \right]^B  $}}\\
 & \hspace{2.5cm} \text{\scalebox{0.9}{$ \displaystyle \times (1-p) \left[ 1- \left( \frac{1-p}{p} \right)  \left( \frac{a}{1-b} \right)^t \right],$}}
\end{split}  \label{eq9}
\end{equation}
whereas $f_n = 0$ for all other $n$.  \label{lemma5}
\end{lemma}

\begin{IEEEproof}
First, by fixing $j=1$ in \eqref{u_n} we observe that $u_{t+1}^V = t p_f^{t-1} (1-p_f)^2 $ for all $t \in \{1,2, \ldots, r_1\}$. Now, recall that $r_1 = r+1$ at $\epsilon_r^+$ whereas $r_1 =r$ at $\epsilon_r^-$. This implies that $u_{t+1}^V (\epsilon_r^+) = u_{t+1}^V (\epsilon_r^-)$ for all $t \leq r$ which corresponds to all agent indices $n \leq r+1 = l_1$. This implies that $\Delta u_n^V = 0$ and therefore $f_n =0$ for all $n \leq l_1$.

Next, consider the agent index $n = l_j +1$ for any fixed $j \in \mathbb{N}$. Here, $u_n^V(\epsilon_r^-) = 0$. Therefore, 
\begin{align}
f_n = (1-p)  u_i^B (\epsilon_r^{+}) \left[ 1 - \Big( \frac{p}{1-p} \Big) \frac{ u_i^G (\epsilon_r^{+})}{u_i^B (\epsilon_r^{+})} \right].  \label{eq5}
\end{align}
Now the ratio $u_i^G (\epsilon_r^{+}) / u_i^B (\epsilon_r^{+})$ can be simplified as follows:
\begin{align}
\frac{u_i^G (\epsilon_r^+)}{u_i^B (\epsilon_r^+)} &= \frac{ \left[ r_j (1-p_f)^2 p_f^{r_j-1} \displaystyle \prod_{i=1}^{j-1} (r_i (1-p_f) p_f^{r_i}) \right]^G}{\left[ r_j (1-p_f)^2 p_f^{r_j-1} \displaystyle \prod_{i=1}^{j-1} (r_i (1-p_f) p_f^{r_i}) \right]^B} \nonumber \\
&= \left( \frac{1-a}{b} \right)^{j+1} \Big( \frac{a}{1-b} \Big)^{\displaystyle \sum_{i=1}^{j} r_i -1} \overset{(a)}{=} \frac{p}{1-p}. \label{eq6}
\end{align} 
Step $(a)$ follows from identities \eqref{identity1}, \eqref{identity2}. Substituting the value obtained in \eqref{eq6} in \eqref{eq5} implies that 
\begin{align}
f_n = 0 \quad \text{at} \;\; n = l_j + 1, \; \text{for every} \;\; j \in \mathbb{N}.  \label{eq11}
\end{align} 

Next, for a fixed $j \in \mathbb{N}$, consider the indices $n = l_j + t +2$ for $ t = 0,1,\ldots, r-1.$ The expression for $u_n$ in \eqref{u_n} yields
{ \small
\begin{align}
u_n (\epsilon_r^+) &= t (1-p_f)^2 p_f^{t-1} \prod_{i=1}^{j} (r_i (1-p_f) p_f^{r_i}),  \;\;\; r_1 = r+1, \label{eq7} \\
u_n (\epsilon_r^-) &= (t+1) (1-p_f)^2 p_f^{t} \prod_{i=1}^{j} (r_i (1-p_f) p_f^{r_i}), \;\;\; r_1 = r. \label{eq8}
\end{align}}%

\noindent This difference in the above expressions is once again due to the different values of $r_1$ for the two cases. Using \eqref{eq7} and \eqref{eq8} to evaluate $\Delta u_n $ gives
{\small
\begin{align}
\Delta u_n = - (1-p_f)^3 \big( r p_f^r (1-p_f) \big)^{j-1} p_f^{r+t} (r-t) \;\; \text{at} \;\; n = l_j +t+2. \label{eq14}
\end{align} }%
Now the expression for $f_n$ in \eqref{f_i} can be rearranged into
\begin{align}
f_n = \Delta u_n^B  (1-p) \left[ 1- \Big( \salpha \Big) \frac{ \Delta u_n^G }{ \Delta u_n^B} \right]. \label{eq10}
\end{align} 
Here, using \eqref{eq14}, the ratio $\Delta u_n^G / \Delta u_n^B$ can be evaluated as
{\small
\begin{align*}
\frac{\Delta u_n^G}{\Delta u_n^B} & = \left( \frac{1-a}{b} \right)^{j+2} \! \Big( \frac{a}{1-b} \Big)^{r j + t } \overset{(b)}{=} \Big( \frac{1-p}{p} \Big)^2 \Big( \frac{a}{1-b} \Big)^t,
\end{align*}}
where Step $(b)$ follows from identities \eqref{identity1}, \eqref{identity2}. Substituting this ratio in \eqref{eq10} gives the required expression for $f_n$ at $n = l_j + t+ 2$ for $t = 0,1,\ldots,r-1,$ as stated in Lemma \ref{lemma5}. Note that at $t = r-1$, $n = l_j + r +1 = l_{j+1}$. Therefore, the index $n = l_j$ for every $j$ is also exhausted by \eqref{eq9} when $t = r-1$. In this manner all agent indices have been covered, thus completing the proof.
\end{IEEEproof}

\renewcommand\IEEEproofname{Proof of Theorem \ref{thm:welfare_agent_n_@eps_r}}

\begin{IEEEproof}
Now, equipped with the above Lemmas, we begin the main proof that shows that the expression in \eqref{d_i_plus_f_i_master} is non-negative for all $n$. First, we evaluate the following term using \eqref{eq9}. 
\begin{align}
\sum_{i=l_j + 1}^{l_{j+1}} f_i & \overset{(a)}{=} \; 0 + \sum_{i=l_j + 2}^{l_{j+1}} f_i, \\
\begin{split} \overset{(b)}\geq - \text{\scalebox{0.9}{$ \displaystyle  \left[ (1-p_f)^3 \big(r p_f^r (1-p_f) \big)^{j-1} p_f^{r} \sum_{t=0}^{r-1} p_f^t (r-t) \right]^B  $}}\\
\text{\scalebox{0.9}{$ \displaystyle \times (1-p) \left[ 1- \left( \frac{1-p}{p} \right)   \right].$}}
\end{split}  \label{eq12}
\end{align}
Step $(a)$ follows from \eqref{eq11}. To obtain the inequality in Step $(b)$, we ignore the term $(a/(1-b))^t$ in the expression for $f_n$ in \eqref{eq9}. Doing this provides a lower bound to each $f_i$ and thereby to the net sum.

Next, for a fixed $j \in \mathbb{N}$, we want to prove the inequality,
\begin{align}
d_{l_j} + \sum_{i=l_j + 1}^{l_{j+1}} \! f_i \geq 0. \label{eq13}
\end{align}
Here, $d_{l_j}$ is given in \eqref{eq0} and the second term can be bounded from below using the inequality in \eqref{eq12}. Thus, a sufficient condition for inequality \eqref{eq13} to hold is,
\begin{align}
&\text{\scalebox{0.9}{$ p \left[ p_f^{r_j} \displaystyle \prod_{i=1}^{j-1} r_i (1-p_f) p_f^{r_i} \right]^B \! \! \! \! \left( \frac{a}{1-b} -1 \right) $}} \geq \nonumber \\
\begin{split}  \text{\scalebox{0.9}{$ \displaystyle  \left[ (1-p_f)^3 \big(r p_f^r (1-p_f) \big)^{j-1} p_f^{r} \sum_{t=0}^{r-1} p_f^t (r-t) \right]^B  $}}\\
\text{\scalebox{0.9}{$ \displaystyle \times (1-p) \left[ 1- \left( \frac{1-p}{p} \right)   \right].$}}
\end{split} 
\end{align}
Here, using the geometric series formula, the summation term can be bounded as \scalebox{0.85}{$ \displaystyle \sum_{t=0}^{r-1} p_f^t (r-t) \leq \frac{r}{1-p_f}$}. This together with cancelling out of the terms common on both sides of the inequality reduces the sufficient condition to 
\begin{align}
\frac{1}{r} \geq (1-a).  \label{eq14}
\end{align} 
Now, as $1/r =\eta$ at each threshold $\epsilon_r$, proving \eqref{eq14} implies showing that $\eta \geq (1-a)$ at every $\epsilon_r$. It can be shown that this is in fact true not just at the thresholds $\{\epsilon_r\}$ but for any $\epsilon \in (0,1)$. Therefore, the inequality suggested in \eqref{eq13} is proved.

Now, as $f_i$ is negative for all $i \in \{l_j +1, \ldots, l_{j+1} \}$ as is evident from the ``minus'' sign in \eqref{eq9}, the inequality in \eqref{eq13} also implies that for any $j \in \mathbb{N}$,
\begin{align}
d_{l_j} + \sum_{i=l_j + 1}^{n} f_i >0 \quad \forall \; n \in \{l_j +1, \ldots, l_{j+1} \}.
\label{less_than0_2}
\end{align}
Now, for any fixed agent index $n$, let $J := \max \{j : l_j \leq n \}$. Then, the summation considered in \eqref{d_i_plus_f_i_master} can be broken into smaller sums as shown below.  
\begin{align}
\sum_{i=1}^{n} (d_i + f_i) &= \sum_{j=1}^{J-1} \Big( d_{l_j} \! + \!\! \sum_{i=l_j + 1}^{l_{j+1}} \! f_i \Big) + \; d_{l_J} \! + \!\! \sum_{i=l_J + 1}^{n} \! f_i \label{sum_dist}\\
& \overset{(c)}{\geq} 0. 
\end{align}
Step $(c)$ follows from the fact that each of the $J$ terms in \eqref{sum_dist} is non-negative as per the inequality \eqref{less_than0_2}.

\end{IEEEproof}

\section{Proof for equation \eqref{eps_lim_expression}}
\label{app:eps_lim_proof}


\begin{IEEEproof}
First, we first obtain the limitng value for the term $p_f^r$ as $r \rightarrow \infty$. For this, recall that as $r = \left\lfloor 1 / \eta \right\rfloor $, we have  $p_f^{1 / \eta} \leq p_f^r < p_f^{1 / \eta -1}$. Since the upper and lower bounds converge to the same limit, it implies that
\begin{align}
\underset{r \rightarrow \infty}{\lim} p_f^r = \underset{\epsilon \rightarrow 1}{\lim} p_f^{1 / \eta}.
\label{p_f^r_limit}
\end{align}
Further, given that $\eta \triangleq \log(\frac{a}{1-b}) / \log(\frac{p}{1-p}) $, the numerator of $\eta$ can be bounded using the inequality: $ 1- x^{-1} \leq  \log(x) \leq x-1$, which holds for all $x > 0$. As a result, $\eta$ can be bounded as
\begin{align}
\frac{(2p-1)(1-\epsilon)}{ \big( p+(1-p)\epsilon \big) \log \alpha} \leq \eta \leq \frac{(2p-1)(1-\epsilon)}{ \big( 1 - p(1-\epsilon) \big) \log \alpha}.
\label{eta_bound}
\end{align}
This in turn extends to the following bounds on $p_f^{1 / \eta}$,
\begin{align}
p_f^{\frac{1}{1-p_f} at } \leq p_f^{1 / \eta} \leq p_f^{\frac{1}{1-p_f} (1-b)t },
\label{p_f^eta_upper_lower}
\end{align}
where \scalebox{0.95}{$ t = \frac{1}{\alpha-1} \log \alpha$} for \scalebox{0.95}{$V=G$}, and \scalebox{0.95}{$ t= \frac{\alpha}{\alpha-1} \log \alpha$} for \scalebox{0.95}{$V=B$}. Now, as both $a$ and $1-b$ tend to $1$ as $\epsilon \rightarrow 1$; both the upper and the lower bounds presented in \eqref{p_f^eta_upper_lower} achieve the same limit. Hence, 
\begin{align}
\underset{r \rightarrow \infty}{\lim} p_f^r \overset{\eqref{p_f^r_limit}}{=} \underset{\epsilon \rightarrow 1}{\lim} p_f^{1 / \eta} = \left( \underset{\epsilon \rightarrow 1}{\lim} p_f^{\frac{1}{1-p_f}} \right)^t \overset{(a)}{=} e^{-t},
\label{lim_p_f^r_final}
\end{align}
where Step $(a)$ follows by reducing the limit term such that it maps to the identity: $\underset{x \rightarrow 0^{+}}{\lim} (1-x)^{\frac{1}{x}} = 1/e $.

Secondly, we obtain the limitng value for the term $r(1- p_f)$ which equals the limit for $(1-p_f)/\eta$ because $r = \left\lfloor 1 / \eta \right\rfloor $. Now, inequality \eqref{eta_bound} leads to the following bounds on $(1-p_f)/\eta$:
\begin{align*}
\frac{(1-b)(1-p_f)}{(2p-1)(1-\epsilon)}  \log \alpha \leq \frac{1-p_f}{\eta} \leq \frac{a(1-p_f)}{(2p-1)(1-\epsilon)} \log \alpha.
\end{align*} 
But as both $a$ and $1-b$ tend to $1$ as $\epsilon \rightarrow 1$; both the upper and the lower bounds presented above achieve the same limit. Hence, 
\begin{align}
\underset{\epsilon \rightarrow 1}{\lim} \; r(1- p_f) = \underset{\epsilon \rightarrow 1}{\lim} \frac{1-p_f}{\eta} = \underset{\epsilon \rightarrow 1}{\lim} \frac{(1-p_f) \log \alpha}{(2p-1)(1-\epsilon)} = t,
\label{r(1-p_f)_lim_final}
\end{align}
where $t$ has been defined for $V \in \{G,B\}$ in \eqref{p_f^eta_upper_lower}.

Finally, we use the limits obtained in \eqref{lim_p_f^r_final} and \eqref{r(1-p_f)_lim_final} for evaluating the limit stated in \eqref{eps_lim_expression}, which gives the desired result.
\end{IEEEproof}

\section{Cascade probabilities and welfare for Scenario $(c)$}
\label{app:scene_c}


In Scenario $(c)$, for the specific value: $\beta = \beta_{\text{sym}} := \frac{\epsilon}{1+\epsilon}$, it follows that  $\epsilon_Y = \epsilon_N = \beta_{\text{sym}}$ in \eqref{epsY_epsN_(c)}. This implies that the channel between $A_i$ and $O_i^{'}$ is effectively a BSC with cross-over probability $\beta_{\text{sym}}$. Note that given $\epsilon\in [0,1)$, $\beta_{\text{sym}} \in [0,0.5)$. The channel is then equivalent to the symmetric noise channel between $A_i$ and $O_i^{'}$ considered in \cite{Tho}. Results in \cite{Tho} show that, for all times $n$ until a cascade occurs, $s_n := n_Y -  n_N $, which is the difference between the number of $Y$'s $(n_Y)$  and $N$'s $(n_N)$ present in $\mathcal{H}_{n}$, is a sufficient statistic of this history. Thus, until a cascade occurs, the update rule for $s_n$ is given by
\begin{align}
s_n = \begin{cases} s_{n-1} + 1 \quad & \text{if} \;\; O_n^{'} = Y,\\
s_{n-1}-1  \quad & \text{if} \;\; O_n^{'} = N.
\end{cases}
\label{mc_update_s_n}
\end{align}
Whereas, once $s_n = k $ $( -k )$, a $Y \,(N)$ cascade begins and $s_n$ stops updating (due to Property \ref{prop1}). For our scenario, integer $k$ in eq. $(6)$ of \cite{Tho} is redefined as $k := \left \lfloor  \log_{ \text{\scalebox{0.9}{$ (1-a_{\text{sym}}) / a_{\text{sym}} $}} } \big( \text{\scalebox{0.87}{$\frac{1-p}{p}$}} \big) \right \rfloor +1,$ where $a_{\text{sym}} $ denotes the probability that observation $O_i^{'}$ follows the true value $V,$ conditioned on $V$ and that agent $i$ follows $S_i$. 
\begin{align}
a_{\text{sym}} &= \mathbb{P} (O_i^{'} = Y \vert V=G) \overset{(a)}{=} \mathbb{P} (O_i^{'} = N \vert V=B), \label{a_sym} \\
&= p(1-\beta_{\text{sym}}) +(1-p)\beta_{\text{sym}}.  \nonumber
\end{align}
Here, Step $(a)$ follows since the channel between $V$ and $O_i^{'}$ is also a BSC, and has $a_{\text{sym}} $ as its cross-over probability.

Thus, $\{s_n\}$ is a random walk over a finite state-space $S = \{-k, \ldots, -1,0,1, \ldots, k\}$, where states $k$ and $-k$ are absorption states that correspond to $Y$ and $N$ cascades, respectively. Further, it follows from \eqref{mc_update_s_n} that conditioned on $V$, $ \{ s_n \}$ starts from state $0$, moves to the right by $1$ w.p. $\mathbb{P}(O_n^{'} = Y \vert V)$ or to the left by $1$ w.p. $\mathbb{P}(O_n^{'} = N \vert V) $ until a cascade occurs, where these probabilities are defined in terms of $a_{\text{sym}}$ in \eqref{a_sym}. As the state-space $S$ for this random walk is finite, absorption probabilities can be obtained by simply solving a system of linear equations. This yields the probability of a $Y$ cascade given $V=G$, denoted by $ ^{(c)}\mathbb{P}_{Y \text{-cas}}^G (\epsilon, \beta_{\text{sym}})$ as 
\begin{align}
^{(c)}\mathbb{P}_{Y \text{-cas}}^G (\epsilon, \beta_{\text{sym}}) = \frac{a_{\text{sym}}^{k}}{(1-a_{\text{sym}})^{k} + a_{\text{sym}}^{k}}.
\label{P_Ycas_scene_c}
\end{align}
Further, the symmetric structure of the binary channel between $V$ and $O_i^{'}$, when agent $i$ follows $S_i$, implies that $ ^{(c)}\mathbb{P}_{Y \text{-cas}}^G (\epsilon,\beta_{\text{sym}}) = ^{(c)}\mathbb{P}_{N \text{-cas}}^B (\epsilon,\beta_{\text{sym}})$. Applying this equality to \eqref{Pi_expression} and using \eqref{P_Ycas_scene_c} to simplify, yields the asymptotic welfare for this scenario, denoted by $^{(c)}\Pi(\epsilon,\beta_{\text{sym}}) $ and given in \eqref{Pi_scene_c}.

\newpage

\section{Additional comparisons of asymptotic welfares with the Platform Co-ordinator}
\label{app:scenario_plots}

\begin{figure}[h!]
\centering
\subfloat[$p=0.55$]{
\includegraphics[trim = 0cm 0mm 0cm 3mm, width=\linewidth, clip]{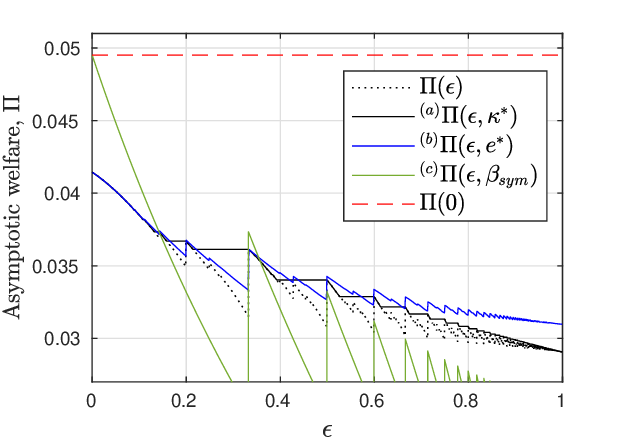}
\label{12a} }
\hfill
\subfloat[$p=0.9$]{
\includegraphics[trim = 0cm 0mm 0cm 3mm, width=\linewidth, clip]{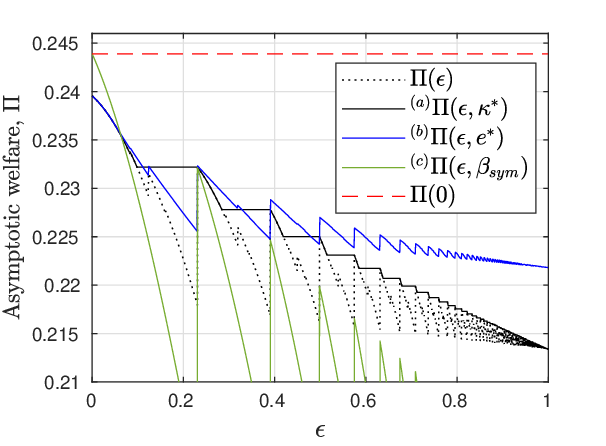}
\label{12b} }
\caption{\small Asymptotic welfare for the indicated scenarios versus the fraction of fake agents given the private signal qualities: $(a)$ $p=0.55$ and $(b)$ $p=0.9$.}
\label{fig:app_welfare_compare}
\end{figure}

In this appendix, we present additional comparisons of asymptotic welfares for the three scenarios with the Co-ordinator, that have been discussed in Section \ref{sec:platform_coord}. In particular, Figures \ref{12a} and \ref{12b} compare the three welfares: $^{(a)}\Pi(\epsilon, \kappa^*)$ , $^{(b)}\Pi(\epsilon, e^*)$ and $^{(c)} \Pi (\epsilon, \beta_{\text{sym}})$ against varying fraction of fake agents $\epsilon$, under private signal qualities $p = 0.55$ and $p = 0.9$, respectively. As done earlier in Figure \ref{fig:welfare_compare} for $p =0.7$, we contrast these welfares with the default welfare in the absence of the Co-ordinator, $\Pi(\epsilon)$ and the baseline welfare $\Pi(0)$. We see that the characteristics of the three scenarios, which were observed for $p=0.7$ (in Fig. \ref{fig:welfare_compare}) are also consistent for the new values of $p$ considered here. Further, for the sequence of Figures: \ref{12a}, \ref{fig:welfare_compare} and \ref{12b}, with increasing values of $p$, we observe that the initial range of $\epsilon$ for which Scenerio $(c)$ performs the best decreases with $p$. Whereas, the range of high $\epsilon$-values at which Scenerio $(b)$ outperforms the rest increases with $p$. Through extensive simulations, we find that the above characteristics of the three scenarios are indeed consistent for all values of $p$. 


\end{document}